\documentclass[a4paper,fleqn,usenatbib]{mnras}

\usepackage{subfig}

\usepackage[T1]{fontenc}
\usepackage{ae,aecompl}
\usepackage{pdflscape}
\usepackage{graphicx, caption}	% Including figure files
\usepackage{amsmath}	% Advanced maths commands
\usepackage{amssymb}	% Extra maths symbols
\usepackage{color}

\usepackage{graphicx, caption}
\def\etal{{et al.~}}

\def\kms{\>{\rm ~km}\,{\rm s}^{-1}}

\def\Msun{\>{\rm M_{\odot}}}

\title[Recoiling SMBHs in analytical and numerical galaxies]{Recoiling supermassive black holes in analytical and numerical galaxy potential}

\author[M. Smole et al.]{
Majda Smole,$^{1}$\thanks{E-mail: msmole@aob.rs, 
micic@aob.rs, 
amitrasinovic@aob.rs}
Miroslav Micic,$^1$ Ana Mitra\v{s}inovi\'c$^1$
\\
$^1$ Astronomical Observatory, Volgina 7, 11060 Belgrade, Serbia}

\date{Accepted XXX. Received YYY; in original form ZZZ}

\pubyear{2017}

\begin{document}
\label{firstpage}
\pagerange{\pageref{firstpage}--\pageref{lastpage}}
\maketitle

\begin{abstract}
We follow trajectories of recoiling supermassive black holes (SMBHs) in analytical and numerical models of galaxy merger remnants
with masses
of $10^{11}\Msun$ and $10^{12}\Msun$.
We construct various merger remnant galaxies in order 
to investigate how the central SMBH mass and the mass ratio of progenitor galaxies
influence escape velocities of recoiling SMBHs.
Our results show that static analytical models of major merger
remnant galaxies overestimate the SMBHs escape velocities.
During major mergers violent relaxation
leads to the decrease of galaxy mass and lower potential at large remnant radii.
This process is not depicted in static analytical potential but clearly seen in our numerical
models. Thus, the evolving numerical model is a more realistic description of dynamical
processes in galaxies with merging SMBHs.
We find that SMBH escape velocities in numerical major merger remnant galaxies can be up to 25 per cent
lower compared to those in analytical models.  
Consequently, SMBHs in numerical models generally reach greater galactocentric distances and spend more time 
on bound orbits outside of the galactic nuclei.
Thus, numerical models predict a greater number of spatially-offset active galactic nuclei (AGNs).
\end{abstract}


%%%%%%%%%%%%%%%%%%%%%%%%%%%%%%%%%%%%%%%%%%%%%%%%%%

%%%%%%%%%%%%%%%%% BODY OF PAPER %%%%%%%%%%%%%%%%%%

\section{Introduction}

According to the hierarchical galaxy formation, 
massive galaxies in local Universe have experienced numerous mergers.
During the merger of two galaxies hosting SMBHs, due to the dynamical friction, SMBHs
sink to the centre of the merger remnant, where they form a binary system.
Further interactions with stars and gas 
carry away energy of the system which
leads to the binary hardening.

In gas poor galaxies, SMBH binaries capture stars on centrophilic orbits and then eject them 
at higher velocities which results in decreasement of the separation 
between SMBHs (\citealt{begel80}; \citealt{ebisuzaki}). However, as binary system ejects
stars, their central density decreases and 
once the loss cone is depleted, decay of SMBH binary is expected to stall at separation of
$\lesssim1$ pc, which is so-cold \textquotedblleft final parsec problem\textquotedblright $~$
(e.g. \citealt{milosavljevic01}).
SMBH binary coalescence timescale in spherical systems overcomes Hubble time and 
represents a bottleneck for SMBH mergers.
On the other hand, numerical simulations have shown that SMBH coalescence timescales
in triaxial and axisymmetric merger remnants are much shorter ($\sim1$Gyr) and SMBH could be brought to separation 
$\ll1$ pc before all stars on intersecting orbits have been ejected (e.g \citealt{Berczik}; \citealt{Berentzen}; 
\citealt{Preto}; \citealt{Khan11}; \citealt{Khan13}; \citealt{Gualandris2017}; \citealt{Rantala}).
After gas rich galaxy merger, gas that is fueled to the central regions of galaxy
may play important role in SMBH binary hardening. Interactions
with massive gaseous nuclear disc and gas clumps can lead to the 
efficient SMBH coalescence within $10^7$ yr (\citealt{escala0405}; \citealt{mayer07};
\citealt{Fiacconi}; \citealt{mayer13}; \citealt{Roskar}; \citealt{delval}; \citealt{Goicovic}; \citealt{Khan16}).
However, in minor merger remnants SMBH sinking timescale depends on 
the orbital parameters and can be much longer \citep{Callegari0911}.
Similar results have been found in the recent study by \cite{Tremmel}
who used {\fontfamily{qcr}\selectfont Romulus25} cosmological simulation
to investigate timescales of SMBHs sinking down to sub-kpc scales in 
galaxies with different properties.
Their results suggest that formation of close SMBH binary is strongly influenced
by galaxy morphology and mass ratio of the progenitor galaxies, while some 
galaxy mergers do not necessary lead to SMBH merger.

Once the separation between SMBHs becomes $\lesssim10^{-3}$ pc,
gravitational wave radiation will
efficiently extract angular momentum and energy from
the binary system, causing rapid SMBHs merger \citep{begel80a}.
Any asymmetry in the binary system, caused by SMBHs with unequal masses and/or
spins, will lead to the asymmetric emission of
gravitational radiation and SMBH recoil.
Gravitational waves propagate in a preferential direction
due to non-zero net linear momentum and the centre
of mass of the binary recoils in the opposite direction
\citep{redmount}.
In this process newly formed SMBH receives a kick, whose magnitude depends on 
the mass ratio of SMBHs, the spin magnitude and orientation with respect to
the binary orbital plane, and the eccentricity of the orbit.
With a breakthrough of the full numerical relativity simulations it has been shown that
kick velocities can be as high as $\sim4000\kms$ (\citealt{gonzalez}; \citealt{campa}; \citealt{Lousto11}),
although super-kicks are considered to be rare and 
most SMBHs are ejected at lower velocities.
Super-kicks are expected to occur for maximally spinning SMBHs, 
whose spin vectors prior to a merger are in the orbital plane and
have opposite directions.
However, in the presence of central gas disc torques from the disc can lead 
to SMBH spin alignment, 
and in that case even fast rotating SMBHs will not get kick velocities
larger than $\sim600\kms$ \citep{blecha16}.

Super-kicks can eject SMBHs even from the massive elliptical galaxies, while 
lower kick velocities can be sufficient to remove SMBHs from the dwarf galaxies, globular
clusters, and high redshift 
haloes, which generally have lower masses (\citealt{merritt}; \citealt{micic2006}; 
\citealt{volonteri2007}; \citealt{schni2007} ; \citealt{sesana}; 
\citealt{volonteri2010}; \citealt{micic2011}).
Gravitational wave recoil can have negative influence on SMBH growth through mergers, 
since ejected SMBHs are less likely to
merge with other SMBHs (e.g. \citealt{haiman}; \citealt{merritt}; \citealt{volonteri2007}).

If a SMBH receives kick with amplitude lower than the escape velocity from the host,
it may be displaced from the nucleus, and make several
passages across the centre of the galaxy, before it sinks
back to the core. 
Time scale for orbit decay
depends on the kick velocity and oscillations can last up to several Gyr
(\citealt{madau04}; \citealt{Gualandris}; \citealt{Komossa08}; \citealt{blecha2008}).
Presence of the gas in the galactic centre can significantly influence 
recoiling SMBHs trajectories and in gas 
rich galaxies SMBHs will spend less time
outside of the galactic nuclei (e.g. \citealt{blecha2008}; \citealt{deve09}; \citealt{blecha11}; \citealt{Guedes}; \citealt{sijacki11}).

Effects of the host halo accretion on the recoiling SMBH trajectories have been studied analytically 
by \cite{smole} and \cite{Choksi}.
Both works have shown that host halo growth significantly influences SMBH escape velocities
since SMBHs can be easily ejected from low mass galaxies at high redshifts, while halo accretion at later 
times makes the ejection harder.

Recoiling SMBHs could be observed as spatially or kinematically offset AGNs
(e.g. \citealt{madau04}; \citealt{loeb07}; \citealt{blecha2008}; \citealt{blecha11}; \citealt{sijacki11}).
Spatially-offset AGNs are characterized by an accreting SMBH displaced from the host galaxy centre.
In the case of kinematically-offset AGN broad line region is ejected together with the SMBH,
while narrow line region stays associated with the galaxy nucleus. Hence, ejection of the SMBH produces a difference
between velocities of these two regions.
Numerous candidate offset AGNs have been observed, although alternative
explanations cannot be ruled out (we refer to \citealt{blecha16} for more detailed summary of offset AGN 
observations).

In order to test the observability of the offset AGN,
\cite{blecha16} investigated trajectories of recoiling SMBHs using 
Illustris cosmological simulations \citep{Vogelsberger}. 
By extracting properties of merging SMBHs and progenitor galaxies directly from the simulation,
authors constructed analytical models of merger remnants galaxies 
consisted of dark matter halo (DMH), bulge and circumnuclear disc.
Further, they assumed different SMBH spin distributions
to estimate resulting kick velocities which are then assigned
to central SMBHs whose trajectories are integrated.

In this paper, we apply similar analytical model as described in \cite{blecha16}
to construct merger remnant galaxies of specific masses. 
Instead of using cosmological simulation to extract properties of 
remnant galaxies, in our model merger remnant galaxy is 
determined by its total mass, the
central SMBH mass and the scaling relations.
In addition to this, we construct numerical models of the merging galaxies
in order to obtain remnants with characteristics similar to those in analytical models.
We calculate escape velocities of SMBHs ejected
from the galaxies which have been formed in major or minor mergers.
The capability of different hosts to retain recoiling SMBHs is explored
with special focus on differences between analytical and numerical 
models of galaxies.
In Section~\ref{method} we describe the method. In Section~\ref{results} we present our results. 
We summarize and discuss our results in Section~\ref{discussion}.

%This is a simple template for authors to write new MNRAS papers.
%See \texttt{mnras\_sample.tex} for a more complex example, and \texttt{mnras\_guide.tex}
%for a full user guide.

%All papers should start with an Introduction section, which sets the work
%in context, cites relevant earlier studies in the field by \citet{Others2013},
%and describes the problem the authors aim to solve \citep[e.g.][]{Author2012}.

\section{Methods}
\label{method}

We follow trajectories of the recoiling SMBHs in analytical and numerical galaxy potential whose 
components are DMH, bulge and disc. 
Total masses of the considered galaxies are $10^{11}\Msun$ and $10^{12}\Msun$.
The chosen masses are of the greatest
significance for studying recoiling SMBHs. More 
massive galaxies have deep gravitational well
and only rare super-kicks could remove SMBHs from their hosts. In less massive galaxies,
kick velocities $\lesssim100\kms$ could lead to the complete SMBH ejection.
Furthermore, the estimated distribution of recoil kick velocities suggests that
kick amplitudes in interval $100\lesssim v_{\rm{kick}}\lesssim1000\kms$
are the most common kicks \citep{blecha16}. 
In order to investigate if SMBH trajectories are sensitive to the
mass ratio of the progenitor galaxies,
we separately model major (1:1) and minor (1:10) mergers.
Assuming that central SMBHs in major merger progenitors have comparable masses,
the only source of asymmetry in the SMBH binary system are SMBH spins. 
Depending on the spin amplitude and the orientation, SMBH kick velocities in major merger remnants 
could range from few tens up to $\sim2000 \kms$. On the other hand, unequal mass SMBHs in 1:10 mergers 
have narrower interval of possible kick velocities, $50\lesssim v_{\rm{kick}}\lesssim500\kms$ \citep{micic11}.

Since dynamical friction force increases with the recoiling SMBH mass, we
% as $\propto M_{\rm{BH}}^2$ (equation \ref{fdf})
consider two different galaxy models. In the first model, merger remnant galaxy
hosts central SMBH with mass ${M_{\mathrm{BH}}}=M_{\mathrm{gal}}/10^5$, 
which is the expected SMBH mass according to $M_{\rm{BH}}-M_{\rm{halo}}$ relation \citep{Ferrarese2002}.
In further text we address this model as galaxies with $M_{\rm{BH}}-M_{\rm{halo}}$ central SMBH.
In the second model, SMBH mass is
${M_{\mathrm{BH}}}=M_{\mathrm{gal}}/10^3$, which is the
maximal expected SMBH mass derived from the following argument:
if mass in baryons is approximately 16 per cent 
of the dark matter (DM) mass and if a typical 
SMBH mass to bulge mass ratio is 0.5 per cent \citep{kh},
then $\sim8\times10^{-4}$  
of the halo mass is the gas that SMBH can accrete. This model will be addressed as galaxies with
over-massive central SMBH.
Since scaling relations suggest that central SMBH mass and properties of the host galaxies are tightly coupled,
an additional motivation to compare escape velocities from galaxies with different central SMBH masses 
arises. With these two models we are able to follow SMBH ejections in galaxies with typical SMBH masses and
also in galaxies with unusually massive central SMBH.
More detailed description of our models is given in Section \ref{prog_galaxy}.
Parameters of our galaxy models are given in Table~\ref{tabela_prog} and Table~\ref{tabela_remnant}.

For each of the galaxy models described above we use the following procedure:
1) generate initial conditions for numerical pre-merger galaxy,  
2) evolve galaxy in isolation in order to test stability of each galaxy component, 
3) produce the exact analytical counterpart,
4) test the accuracy of our numerical method by comparing,
mass profiles, escape velocities and recoiling SMBH trajectories in numerical and analytical pre-merger galaxies,
5) simulate galaxy merger,
6) compare SMBH trajectories in numerical merger remnants to those in analytical
models of isolated galaxies.
Initial conditions for numerical models are generated using 
{\fontfamily{qcr}\selectfont GalactICS} code (\citealt{Kuijken}; \citealt{widrow05}; \citealt{widrow08}),
while galaxy evolution, mergers and trajectories of kicked SMBHs are simulated using {\fontfamily{qcr}\selectfont
GADGET-2} code \citep{springel2}.
Our main goal is to test how redistribution of mass within post-merger galaxy
affects its capability to retain a recoiling SMBH.
We describe each of these steps in more details bellow.

\subsection{Progenitor galaxy models}
\label{prog_galaxy}

\subsubsection{Numerical models}

Initial conditions for numerical models are generated using 
{\fontfamily{qcr}\selectfont GalactICS} code (\citealt{Kuijken}; \citealt{widrow05}; \citealt{widrow08}).
For the given SMBH mass and the total galaxy mass, masses of individual galaxy components are calculated from scaling 
relations.
Bulge mass is derived using $M_{\mathrm{bulge}}-M_{\mathrm{BH}}$ relation \citep{marconi-hunt}, while 
disc mass is estimated using 
$M_{\mathrm{DM}}-M_{\mathrm{gal,b}}$ relation \citep{moster}.
$M_{\mathrm{gal,b}}$ represents total galaxy mass in baryons, i.e. total disc and bulge mass, thus the disc mass
can be calculated subtracting the bulge mass from the total mass in baryons, estimated from the relation.
$M_{\mathrm{bulge}}-M_{\mathrm{BH}}$ relation predicts that galaxies that harbor over-massive central SMBHs also have massive bulges,
so the total baryonic component of the galaxy estimated using $M_{\mathrm{DM}}-M_{\mathrm{gal,b}}$
relation is in the form of bulge, and in those models disc component is neglected.

Initial conditions for gas particles are calculated from the previously generated N-body
models, following the procedure given by \cite{gajda}. We randomly select 20 per cent 
of disc particles and convert them to gas particles with initial temperature set to
$2000~\mathrm{K}$. Gas particle velocities are calculated as the tangential velocities
at the galaxy mid-plane, while radial and vertical velocities are set to zero.
Gas component is only included in simulations of galaxies whose central SMBH has mass
of ${M_{\mathrm{BH}}}=M_{\mathrm{gal}}/10^5$. Galaxies with over-massive SMBHs do not have disc or gas components and we will refer to
those galaxies as elliptical.

Each progenitor galaxy is represented with $N=10^6$ particles. While distribution of
particles between galaxy components, as well as particle masses, depends on the galaxy model. For instance, 
particle distribution in major merger progenitor of $10^{12}\Msun$
galaxy which harbors $M_{\rm{BH}}-M_{\rm{halo}}$ SMBH
is $N_{\rm{halo}}\sim5\times10^5$, $N_{\rm{disc}}\sim3.6\times10^5$, $N_{\rm{bulge}}\sim5\times10^4$ and $N_{\rm{gas}}\sim9\times10^4$,
while major progenitor of galaxy with over-massive central SMBH 
has different distribution, $N_{\rm{halo}}\sim6\times10^5$ and $N_{\rm{bulge}}\sim4\times10^5$.
With this mass resolution, SMBH mass is always at least one order of magnitude larger
than a DM particle mass, i.e. two orders of magnitude larger than a stellar or gaseous particle mass.
Since we are only interested in general interaction between ejected SMBH and the host galaxy,
this mass ratio is sufficient to track the dynamic of the recoiling SMBH.
SMBH is represented as one particle, placed at the centre of mass calculated with
respect to the most gravitationally bound 5 per cent bulge particles.

A fixed value for gravitational softening length parameter, $\epsilon=0.1~\mathrm{kpc}$, is used in all simulations and 
for all particle types.
Value of the softening length parameter scales with the number of particles in the system  and with
dimension of the system  as
$R/N^{1/3}<\epsilon<R/N^{1/2}$ \citep{binney}. In practice, 
several criteria on the selection of the optimal softening length have been proposed (e.g. \citealt{merritt96};
 \citealt{dehnen}; \citealt{power}; \citealt{zhang}).
Choosing $\epsilon_{\rm{opt}}\simeq2R/N^{1/2}$ as an optimal softening length \citep{zhang}
yields values in range $0.2~\mathrm{kpc}\lesssim\epsilon_{\rm{opt}}\lesssim0.5~\mathrm{kpc}$ for DM particles in different
models, and $\epsilon_{\mathrm{opt}}\lesssim0.1~\mathrm{kpc}$ for baryonic particles.
However, using the same softening length for all particle types significantly reduces
computational time. 
In order to validate the usage of a smaller rather than optimal value of the softening length for DM particles,
we run test simulations adjusting $\epsilon_\mathrm{DM}=0.5~\mathrm{kpc}$ for DM particles and 
$\epsilon_\mathrm{b}=0.1~\mathrm{kpc}$ for baryonic particles.
Test runs did not show any differences in galaxy mass profiles over 3 Gyr compared to the runs with a fixed value
of gravitational softening length, $\epsilon=0.1~\mathrm{kpc}$.
Additionally, difference between SMBH escape velocity, which is calculated over 10 Gyr, is bellow 1 per cent.

\begin{figure*}

\begin{minipage}{170mm}

	\includegraphics[width=\columnwidth]{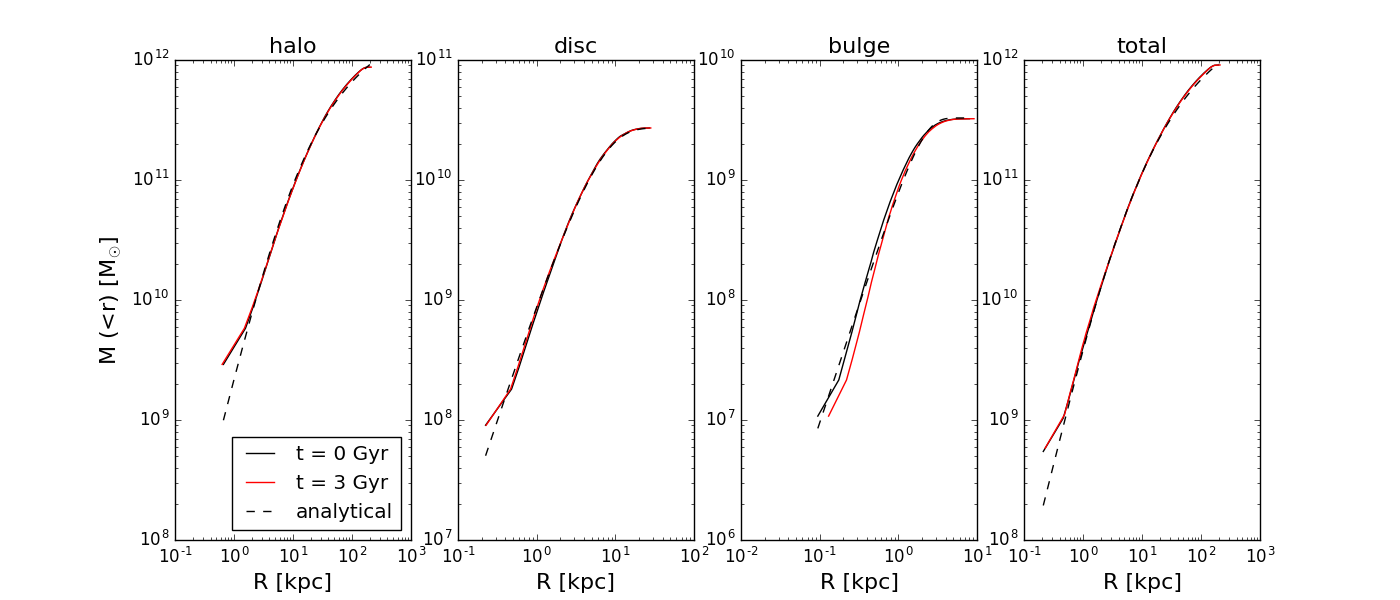}
	\caption{Mass profiles in analytical (dashed lines) and numerical (solid lines) galaxy models.
	For numerical models mass profiles of individual galaxy components are shown at the 
	beginning of the simulation and 3 Gyr later.
	}
    \label{profil_mase_isolation}
    \end{minipage}

\end{figure*}

\begin{table*}
\begin{minipage}{120mm}
\caption{Parameters of progenitor galaxies.} 
\hspace{-2cm}
\label{tabela_prog}
\begin{tabular}{@{}lllllllllllllll}
\hline
$M_\mathrm{gal}$ & $M_\mathrm{BH}$  &merger &progenitor &$M_\mathrm{halo}$&$r_\mathrm{s}$&$c$&$M_\mathrm{bulge}$&$r_\mathrm{bulge}$&$\alpha$ &$M_\mathrm{disc}$&$R_\mathrm{disc} $\\
$[\Msun] $& $[\Msun] $  & ratio& galaxy&$[\Msun] $& $[\mathrm{kpc}]$&&$[\Msun]$&$[\mathrm{kpc}]$ &&$[\Msun]$&$[\mathrm{kpc}]$\\

 \hline

&&1:1&&$4.77\cdot10^{11}$&10.42&15.80&$1.73\cdot10^{9}$&1.90&1.02&$1.58\cdot10^{10}$&3.70\\
&$1\cdot10^{7}$&1:10&primary&$8.72\cdot10^{11}$&9.20&19.50&$3.24\cdot10^{9}$&1.95&1.03&$3.39\cdot10^{10}$&3.60\\
$1\cdot10^{12}$&&&secondary&$8.91\cdot10^{10}$&10.50&10.30&$3.25\cdot10^{8}$&1.72&1.13&$3.49\cdot10^{9}$&2.40\\
 \cline{3-12}

&&1:1&&$2.71\cdot10^{11}$&11.10&11.73&$1.52\cdot10^{11}$&18.70&1.21&/&/\\
&$1\cdot10^{9}$&1:10&primary&$4.97\cdot10^{11}$&12.20&11.50&$4.03\cdot10^{11}$&19.50&1.20&/&/\\
&&&secondary&$5.84\cdot10^{10}$&10.80&9.00&$4.18\cdot10^{10}$&11.00&1.20&/&/\\
 \cline{2-12}
&&1:1&&$4.93\cdot10^{10}$&10.90&8.00&$1.73\cdot10^{8}$&1.50&1.02&$3.80\cdot10^{8}$&2.10\\
&$1\cdot10^{6}$&1:10&primary&$8.94\cdot10^{10}$&10.70&10.50&$3.03\cdot10^{8}$&1.60&1.02&$6.79\cdot10^{8}$&2.50\\
$1\cdot10^{11}$&&&secondary&$8.99\cdot10^{9}$&3.02&11.02&$3.03\cdot10^{7}$&0.26&0.80&$8.64\cdot10^{7}$&2.30\\
 \cline{3-12}

&&1:1&&$2.90\cdot10^{10}$&11.00&6.90&$2.04\cdot10^{10}$&4.60&1.20&/&/\\
&$1\cdot10^{8}$&1:10&primary&$5.32\cdot10^{10}$&7.58&8.80&$3.70\cdot10^{10}$&7.40&1.21&/&/\\
&&&secondary&$5.43\cdot10^{9}$&3.00&11.50&$3.70\cdot10^{9}$&4.10&1.20&/&/\\
\hline
\end{tabular}
\end{minipage}
\end{table*}

After generating initial conditions for numerical galaxy models, we test their stability.
Evolution of each progenitor galaxy is simulated in isolation, 
so that eventual changes in galaxy profiles that emerge due to galaxy model itself can be distinguished
from galaxy interactions.
Galaxy is assumed to be stable if profiles of galaxy components do not significantly 
vary with time over 3 Gyr. Fig.~\ref{profil_mase_isolation} shows an example of stability test run for 
the progenitor of the minor merger galaxy with mass $10^{12}\Msun$ and SMBH mass of $10^{7}\Msun$.
Results of stability tests for other progenitor galaxies are given in Appendix~\ref{A}.

\subsubsection{Analytical models}
\label{analiticki_progenitor}

Further, we make analytical models of progenitor galaxies with the same characteristics as
their numerical counterparts. Density profiles of each of the galaxy components are described bellow
and profile parameters are given in Table~\ref{tabela_prog}.

DMH is described by NFW density profile \citep{nfw}:
\begin{equation} 
\rho_\mathrm{halo}(r)=\frac{\rho_\mathrm{halo,0}}{\frac{r}{r_{\mathrm{s}}}\left(1+\frac{r}{r_{\mathrm{s}}}\right)^{2}},
\end{equation}
where $r_{\mathrm{s}}$ and $\rho_\mathrm{halo,0}$ are the scale radius and the characteristic density, respectively.

Bulge component is modeled as a power-law density spherical potential with an exponential cut-off \citep{bovy13}:
\begin{equation}
\rho_\mathrm{bulge}(r)=\rho_\mathrm{bulge,0}\left(\frac{r_\mathrm{bulge}}{r}\right)^{\alpha}\times \exp[-(r/r_\mathrm{bulge})^2],
\end{equation}
where $r_{\mathrm{bulge}}$ and $\rho_\mathrm{bulge,0}$
represent characteristic bulge radius and density, respectively.

Disc mass distribution is represented as an exponential disc with a surface density:
\begin{equation} 
\Sigma_\mathrm{disc}(R)=\Sigma_\mathrm{disc,0}\times\exp(-R/R_\mathrm{disc}).
\end{equation}
where $\Sigma_\mathrm{disc,0}$ is the central surface density,
$R_\mathrm{disc}$ is disc scale length and $R$ is the distance from the galaxy centre in the disc plane. 
As discussed above, disc component is not included 
in models with over-massive SMBHs, since in those models total baryonic component 
is in the form of massive bulge. 

Table~\ref{tabela_prog} summarize parameters of progenitor galaxies used in analytical models. Those parameters
represent a fit to the mass profiles of galaxy components resulting from numerical models. An example of 
mass profiles of analytical and numerical galaxies is shown on Fig~\ref{profil_mase_isolation}. 
Comparison of mass profiles in analytical and numerical models for all progenitor galaxies is 
shown in Appendix~\ref{A}, together with the stability test of numerical models.

\subsection{SMBH trajectories}
\label{BH_trajectories}

\subsubsection{Analytical models}
 
Next step is to integrate the trajectory of recoiling
SMBH under the influence of potential described above. 
SMBH at a position $\mathbf{x}$ experiences gravitational force 
generated by a distribution of mass $\rho(\mathbf{x'})$ \citep{binney}:
\begin{equation}
\mathbf{F}(\mathbf{x})=M_\mathrm{BH} \mathbf{g}(\mathbf{x}),
\end{equation}
where $\mathbf{g}(\mathbf{x})=G\int\mathrm{d^{3}}\mathbf{x'}\frac{\mathbf{x'}-\mathbf{x}}{|\mathbf{x'}-\mathbf{x}|^{3}}\rho(\mathbf{x'})$
is gravitational field.
If the gravitational potential is defined as:
\begin{equation}
 \Phi (\mathbf{x})\equiv -G\int\mathrm{d^{3}}\mathbf{x'} \frac{\rho(\mathbf{x'})}{|\mathbf{x'}-\mathbf{x}|},
\end{equation}
gravitational field can be calculated as:
\begin{equation}
 \mathbf{g}(\mathbf{x})=-\nabla_{\mathbf{x}} \Phi.
\end{equation}
Here, $\Phi$ is the total gravitational potential from all galaxy components.
However, potential of an exponential disc cannot be calculated analytically. 
We adopt a toy model described by \cite{geehan} where disc mass is 
spherically distributed and whose potential is then:
\begin{equation}
 \Phi_{d}(r)=-2\pi G \Sigma_\mathrm{d,0}R_\mathrm{d}^{2}\left(\frac{1-e^{r/R_\mathrm{d}}}{r}\right).
\end{equation}
The authors have demonstrated that rotation curve for disc component
shows only minor changes when calculated for spherical and axisymmetric case (figure 2, \citealt{geehan}).

In addition to gravitational force, SMBH also experiences dynamical friction force against stars and DM.
We include both stellar and DM dynamical friction via the Chandrasekhar formula \citep{chandra}.
Assuming a Maxwellian velocity distribution, a SMBH moving with velocity $v_\mathrm{BH}$
through a background density $\rho_\mathrm{bulge}$ or $\rho_{\mathrm{halo}}$
will experience dynamical friction force:
\begin{equation} \label{fdf}
\mathrm{\mathbf{f}}_\mathrm{df}=-I(\mathcal{M})\times\frac{4\pi\rho(GM_\mathrm{BH})^2}{\sigma^2}\frac{\mathbfit{v}_\mathrm{BH}}{v_\mathrm{BH}},
\end{equation}
with 
\begin{equation} \label{fdf1}
I(\mathcal{M})=\frac{\mathrm{ln}(\Lambda)}{\mathcal{M}^2}\left(\mathrm{erf}\left(\frac{\mathcal{M}}{\sqrt2}\right)-\sqrt\frac{2}{\pi}\mathcal{M}e^{-\mathcal{M}^2/2}\right),
\end{equation}
where $\mathcal{M}\equiv v_\mathrm{BH}/\sigma$ is the Mach number and Coulomb logarithm is $\mathrm{ln}(\Lambda)=3.1$
\citep{blecha2008}.
Velocity dispersion for stellar component can be calculated using
$M_{\mathrm{BH}}-\sigma_\mathrm{bulge}$ \citep{mcma}  relation.  Following
\cite{tanaka} and \cite{Choksi}  we use simplified model 
and assume constant DM velocity dispersion. Velocity dispersion is
independent of the radius for an isothermal sphere.
Several studies show a tight correlation
between stellar and DM
velocity dispersions, suggesting nearly isothermal total density profiles of quiescent
galaxies (\citealt{Schechter}, \citealt{zahid16}, \citealt{zahid}). 
Here we adopt a relation between DMH mass and 
DM velocity dispersion ($M_{\mathrm{halo}}-\sigma_{\mathrm{DM}}$) derived using Illustris cosmological simulation
\citep{zahid1}.

Trajectory of the recoiling SMBH is governed
by the equation of motion in the form:
\begin{equation}
\ddot{x}=(-\nabla_{\mathbf{x}} \Phi+\mathbfit{a}_\mathrm{{df,DM}}+\mathbfit{a}_\mathrm{{df,bulge}})\mathbf{\hat{x}}.
\end{equation}
SMBH is initially placed at the galaxy centre with an assigned kick velocity in 
orthogonal direction with respect to the disc plane. 
SMBH trajectory is numerically integrated with a time-step $\Delta t=10^{4}$ yr,
using leapfrog integration.

We assign various kick velocities to the SMBHs in our models and follow their 
trajectories until the recoiling SMBH returns to the galaxy halo or until integration
reaches a Hubble time.
Galaxy escape velocity is 
defined as a SMBH kick velocity necessary for a SMBH to return to its host halo after
$\gtrsim10$ Gyr.

\begin{figure}
	\includegraphics[width=\columnwidth]{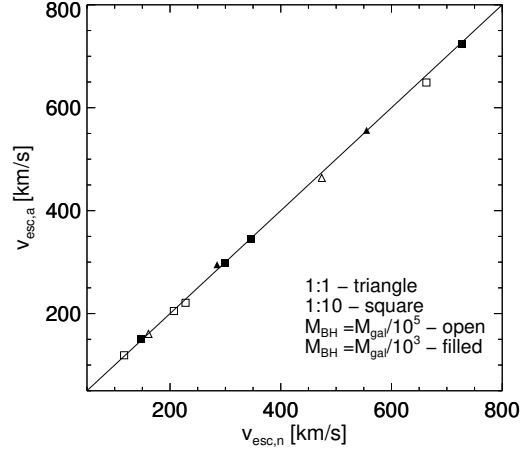}
	\caption{Escape velocities from progenitor galaxies in analytical models as a function of their escape velocities
   in numerical models, for galaxies with different central SMBH masses (open and filled symbols).
Major merger remnants are represented with triangles and minor merger remnants with squares.
}
    \label{vesc_prog}
\end{figure}

\begin{figure}
	\includegraphics[width=\columnwidth]{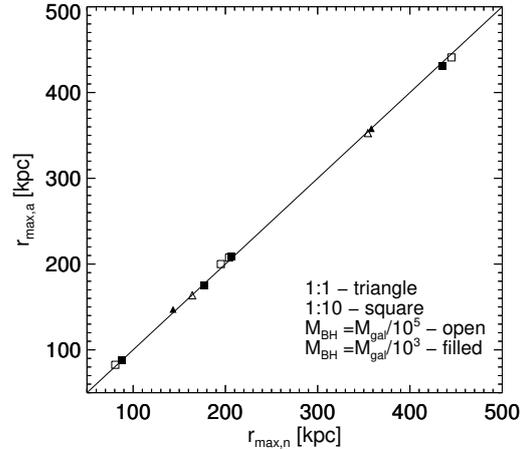}
	\caption{Maximal separation of a recoiling SMBH from the galaxy centre over a Hubble time in analytical models,
	as a function of $r_{\max}$ in numerical models. SMBH kick velocity is equal to the escape velocity 
	for each galaxy model. Notations are the same as in Fig.~\ref{vesc_prog}.}
    \label{rmax_prog}
\end{figure}

\begin{table*}
\begin{minipage}{120mm}
\caption{Parameters of analytical merger remnant galaxies.} 
\hspace{-2cm}
\label{tabela_remnant}
\begin{tabular}{@{}llllllllllll}
\hline
$M_\mathrm{gal}$ & $M_\mathrm{BH}$  &merger &$M_\mathrm{halo}$&$r_\mathrm{s}$&$c$&$M_\mathrm{bulge}$&$r_\mathrm{b}$&$\alpha$ &$M_\mathrm{disc}$&$r_\mathrm{s} $\\
$[\Msun] $& $[\Msun] $  & ratio& $[\Msun] $& $[\mathrm{kpc}]$&&$[\Msun]$&$[\mathrm{kpc}]$ &&$[\Msun]$&$[\mathrm{kpc}]$\\

 \hline

&$1\cdot10^{7}$&1:1&$9.54\cdot10^{11}$&10.70&17.90&$3.51\cdot10^{10}$&6.90&1.13&/&/\\
$1\cdot10^{12}$&&1:10&$9.59\cdot10^{11}$&9.50&19.50&$3.57\cdot10^{9}$&2.10&0.90&$3.80\cdot10^{10}$&3.80\\
 \cline{3-11}

&$1\cdot10^{9}$&1:1&$5.44\cdot10^{11}$&11.18&13.04&$4.44\cdot10^{11}$&18.20&1.10&/&/\\
&&1:10&$5.53\cdot10^{11}$&11.40&14.00&$4.41\cdot10^{11}$&19.50&1.20&/&/\\
 \cline{2-11}
&$1\cdot10^{6}$&1:1&$9.86\cdot10^{10}$&10.10&10.30&$1.32\cdot10^{9}$&3.80&1.11&/&/\\
$1\cdot10^{12}$&&1:10&$9.81\cdot10^{10}$&10.30&10.10&$3.35\cdot10^{8}$&1.50&0.90&$9.48\cdot10^{8}$&2.50\\
 \cline{3-11}

&$1\cdot10^{8}$&1:1&$5.45\cdot10^{10}$&7.50&8.60&$4.45\cdot10^{10}$&7.80&1.10&/&/\\
&&1:10&$5.86\cdot10^{10}$&7.58&8.60&$4.13\cdot10^{10}$&7.40&1.21&/&/\\
\hline
\end{tabular}
\end{minipage}
\end{table*}

\subsubsection{Numerical models}

In the numerical models, SMBH is represented as one massive particle placed at the galaxy centre.
Initial SMBH position is calculated with respect to the most gravitationally bound 5 per cent
bulge particles. As in the analytical models, galaxy escape velocity is defined 
as a SMBH kick velocity necessary for a SMBH to return to its host halo after $\gtrsim10$ Gyr.

In order to validate the accuracy of numerical method, we compare escape velocities from analytical 
and numerical models of progenitor galaxies, as well as the maximal separation from the galaxy centre
a SMBH can reach over 10 Gyr. Results are shown in Fig.~\ref{vesc_prog} and Fig.~\ref{rmax_prog}
for all progenitors of $10^{11}\Msun$ and $10^{12}\Msun$ galaxies.
Escape velocities and maximal separations of SMBHs in analytical and numerical models are in good agreement for all pre-merger
galaxies.

\subsection{Galaxy mergers}

\subsubsection{Numerical models}
\label{mergers}

We use the same orbital parameters for all galaxy merger simulations.
Orbital eccentricity is set to $e=1$, since
N-body cosmological simulations of merging DMHs suggest that 
almost half of all orbits are close to parabolic, with $e=1\pm0.1$
( \citealt{benson}; \citealt{Khochfar}).
The initial separation between galaxies is the sum of their virial radii, 
while pericentric distance is $R_\mathrm{peri}=0.5\%R_\mathrm{vir,1}$,
where $R_\mathrm{vir,1}$ is the virial radius of the primary, i.e. more massive
galaxy. Such a small pericentric distance leads to almost head-on collision
and fast merging process that saves computational time.

SMBHs are added at the beginning of the merger simulations as massive particles, 
placed at the centre of mass of each progenitor galaxy. The centre of mass is calculated 
with respect to the most gravitationally bound 5 per cent
bulge particles.
Separation between galaxies during merger is calculated by tracing
SMBH positions. 
We follow galaxy merger and the SMBH inspiral until separation between SMBHs
drops below the gravitational softening length, i.e. below 0.1 kpc.
Since final stages of SMBH mergers occur on sub-resolution scales, 
we adopt the the similar approach as \cite{blecha16},
and add a time delay for SMBH mergers.
The time delay
depends on SMBH mass ratio $q$ as $0.1~\rm{Gyr}/q$. Here 
we assume that SMBH mass ratio
scales to the mass ratio of
merging galaxies, i.e. $q=1$ for major mergers and
$q=0.1$ for minor mergers.
We follow galaxy mergers until two criteria are met: 1) separation
between SMBHs drops bellow 0.1 kpc, and 2) apocentric
distance for further passages is  $\lesssim10$ kpc. We assume that
the SMBH merger is completed 0.1 Gyr after these criteria are
met in the case of major mergers. Similarly, we add 1 Gyr
time delay in the case of minor mergers.

Using {\fontfamily{qcr}\selectfont Romulus25} cosmological simulation 
\cite{Tremmel} investigated the distribution of time that components of
SMBH binaries spend on separations $\lesssim10$ kpc from each other. Their results suggest that 
most SMBH binaries will form close pairs in less than 1 Gyr, but that 
required time strongly depends on galaxy morphology and mass
ratio of merging galaxies, where some galaxy mergers do not lead to SMBH pair formation.
Close binary formation timescale is short for major mergers of massive galaxies, 
while after minor mergers, where satellite galaxy has low density stellar core,
close pair will form several Gyr after the galaxy merger. 
This is consistent with our results.
We note that in the simulation of minor merger remnant with mass
$M_{\mathrm{gal}}=10^{11}\Msun$ and $M_{\rm{BH}}-M_{\rm{halo}}$
 SMBH, SMBH merger may not be completed within 10 Gyr.
Even though during pericentric passages distance between SMBHs drops below
0.1 kpc, apocentric distances are $\lesssim20$ kpc.
It is unclear whether SMBHs in galaxies with the same orbital parameters would 
merge within a Hubble time, even in gas rich galaxies. 
Merger simulations of gas rich galaxies with the same mass ratio 
suggest that central SMBHs may never merge if their 
orbits have small first pericentric distance, as in our simulations.
During the first pericentric passage strong tidal forces remove the gas from the satellite galaxy,
and the central SMBH stays in the halo of the primary galaxy. At that separation,
the estimated timescale for orbital sinking of the
secondary SMBH may be longer than a Hubble time
\citep{Callegari0911}.
Despite this uncertainty we include this case in our analysis with assumption 
that SMBHs merge after separations between SMBHs drops below our softening length,
and with a relaxed criterion that apocentric
distance for further passages is  $\lesssim20$ kpc.

\subsubsection{Analytical models}

Analytical merger remnant galaxies are modeled in the similar manner as the progenitor galaxies, 
with the same density profiles described in Section~\ref{analiticki_progenitor}.
We assume that major merger of two disc galaxies produces an elliptical galaxy,
while disc component is conserved in minor mergers.
We make numerical models of isolated galaxies and then 
fit their mass profiles in order to produce analytical galaxies 
with the same properties. 
Mass of each galaxy component 
is the sum of DM, stellar and gas masses in progenitor galaxies.
Parameters of analytical post-merger galaxies are
summarized in Table~\ref{tabela_remnant}.

We repeat the same procedure described in Section~\ref{BH_trajectories} in order to follow
trajectories of the recoiling SMBHs in numerical and analytical merger remnant galaxies.

We note that our analytical merger remnant models are spherically symmetric,
unlike numerical models which are not perfectly spherical. 
We perform test runs 
in order to investigate if deviation from spherical symmetry in numerical post-merger galaxies 
could contribute to the difference between models.
Test runs follow a recoiling SMBH in a major merger remnant, since mergers
of equal mass galaxies are expected to produce remnants with higher degree of
asymmetry. However, 
changing direction of a SMBH kick from orthogonal to parallel 
with respect to the disc plane results in less than 1 per cent change in the 
escape velocity.

\section{Results}
\label{results}
%%%%%%%%%%%%%%%%%%%%%%%%%%%%%%%%%%%%%%%%%%%%%%%%%%%%%%%%%%%%%%%%%%%%plot1

Fig.~\ref{vesc} shows galaxy escape velocity in analytical models as a function of escape velocity
   in numerical models for galaxies with mass of $M_{\mathrm{gal}}=10^{12}\Msun$ (upper panel) and
   $M_{\mathrm{gal}}=10^{11}\Msun$ (lower panel). Triangles and squares represent 
   major and minor merger remnants, respectively.
Filled symbols show galaxy models with over-massive central SMBH and open symbols galaxies with a 
SMBH on $M_{\rm{BH}}-M_{\rm{halo}}$ relation.

%%%%%%%%%%%%%%%%%%%%%%%%%%%%%%%%%%%%%%%%%%%%%%%%

\subsection{Comparison between analytical models}
\label{an_vs_an}

\begin{figure}
	\includegraphics[width=\columnwidth]{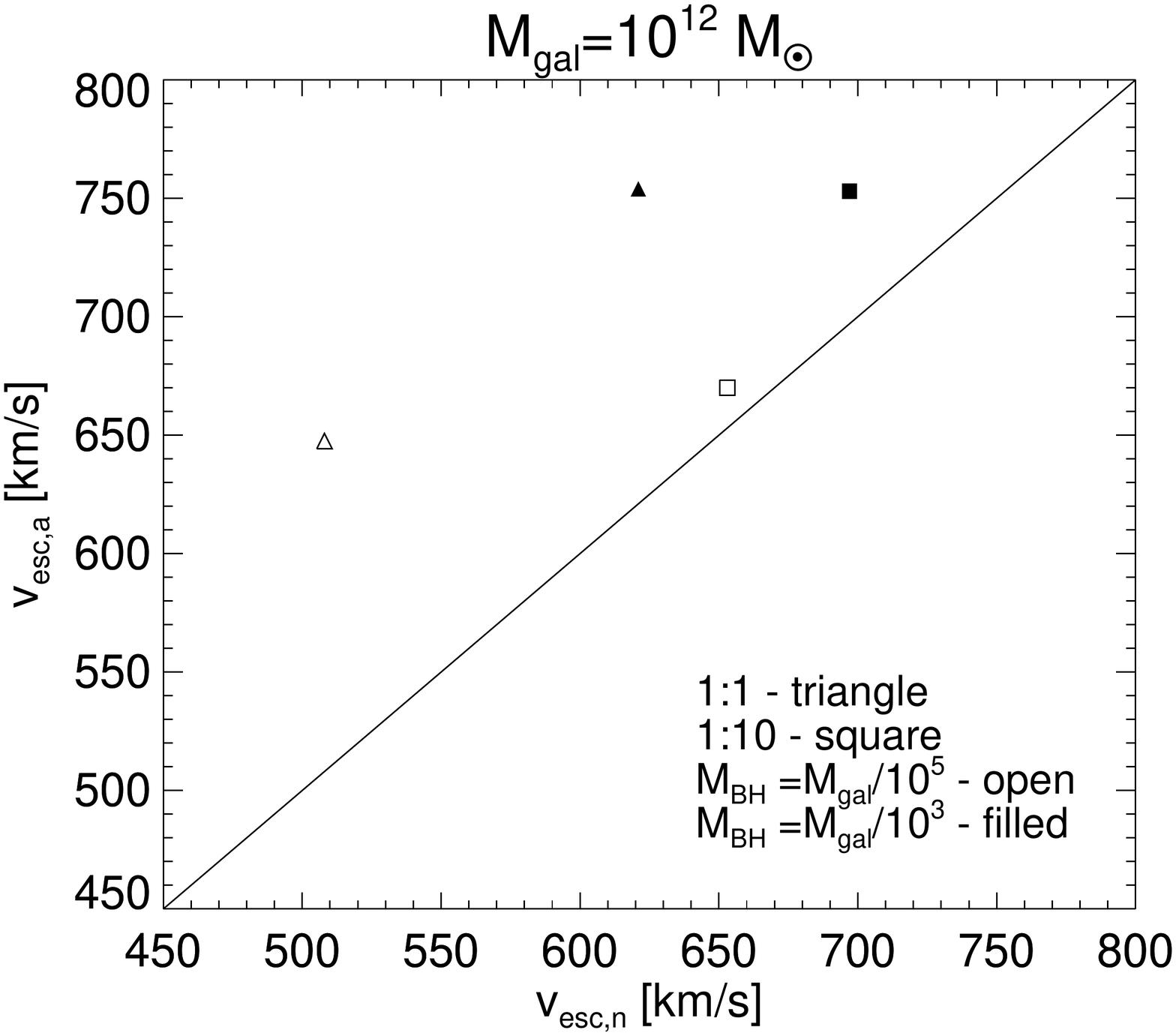}
		\par
	\vspace{0.6cm}
		\includegraphics[width=\columnwidth]{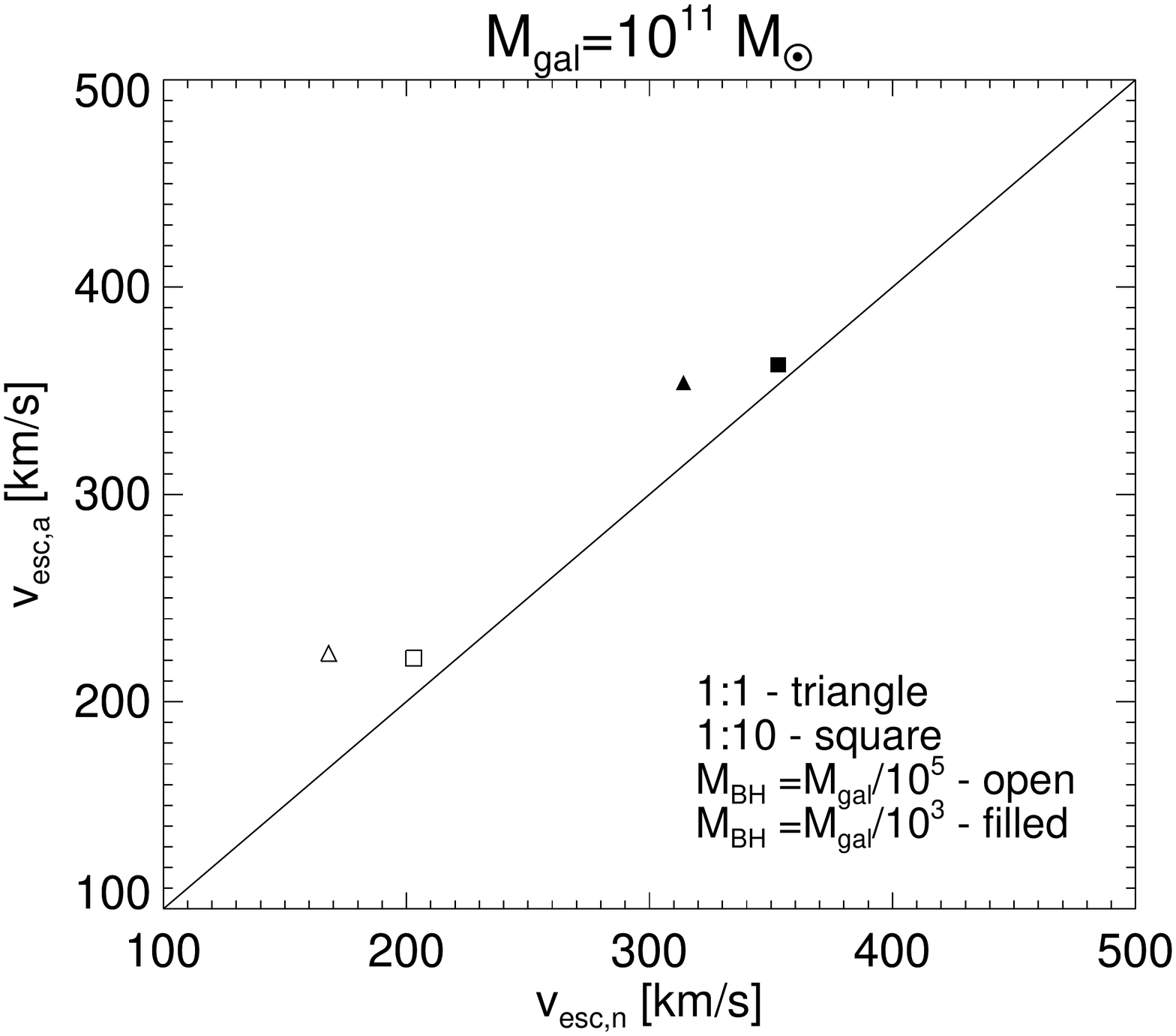}

    \caption{Escape velocities from galaxies in analytical models as a function of their escape velocities
   in numerical models, for galaxies with different central SMBH masses (open and filled symbols).
Major merger remnants are represented with triangles and minor merger remnants with squares.
We note that a relaxed criteria is imposed for minor merger of  $M_{\mathrm{gal}}=10^{11}\Msun$
galaxy with ${M_{\mathrm{BH}}}=M_{\mathrm{gal}}/10^5$ SMBH: we assume that SMBH merger is completed 
1 Gyr after separation between SMBHs drops below 0.1 kpc, while apocentric distance
for further passages is $\lesssim20$ kpc (we refer to Section~\ref{mergers} for more details)}.

    \label{vesc}
\end{figure}

\begin{figure*}
 \begin{minipage}{150mm}

	\centering
	\subfloat[Total mass profile of $10^{12}\Msun$ galaxies]{\includegraphics[width=1\columnwidth]{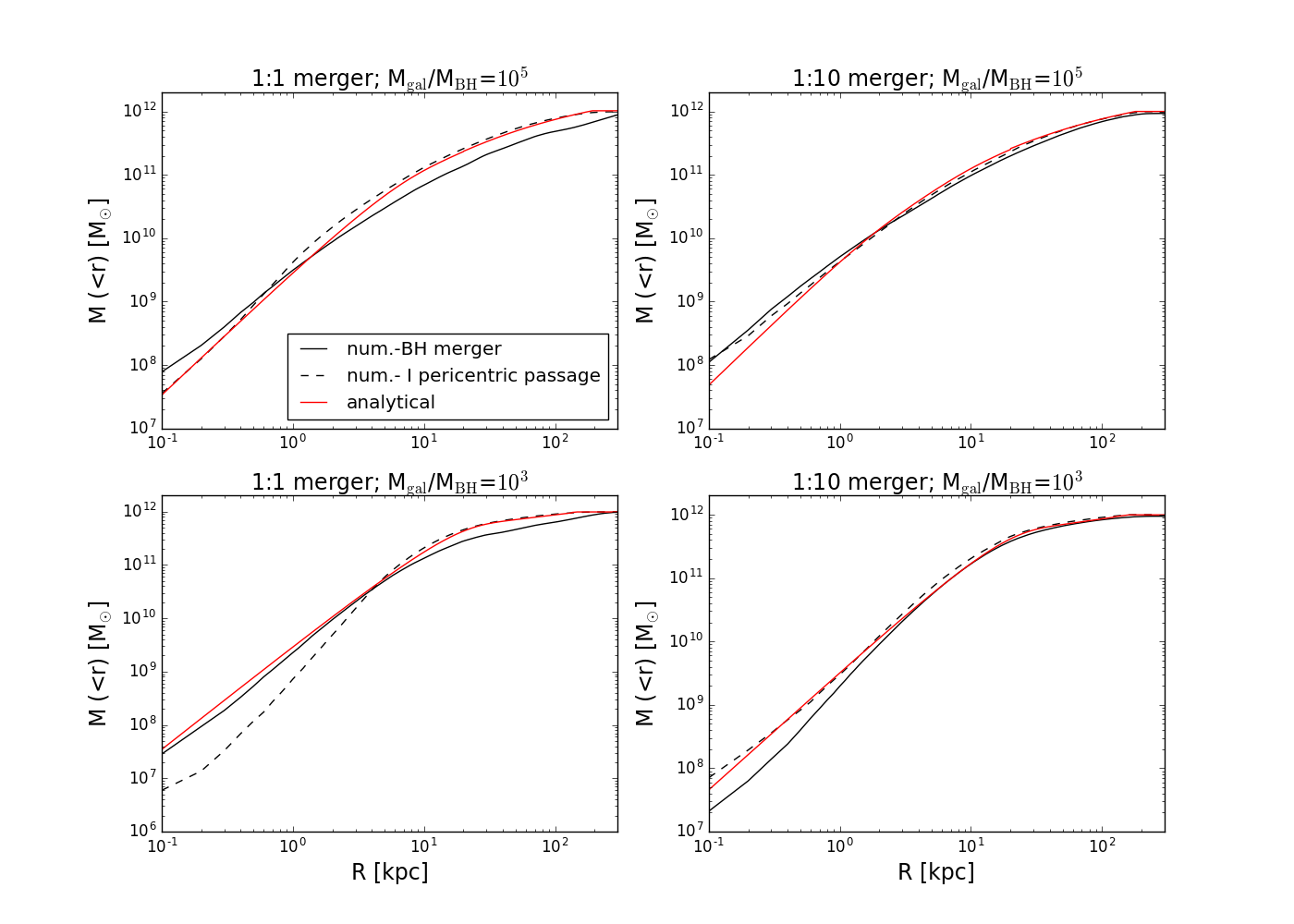}}\\
	\subfloat[Total mass profile of $10^{11}\Msun$ galaxies]{\includegraphics[width=1\columnwidth]{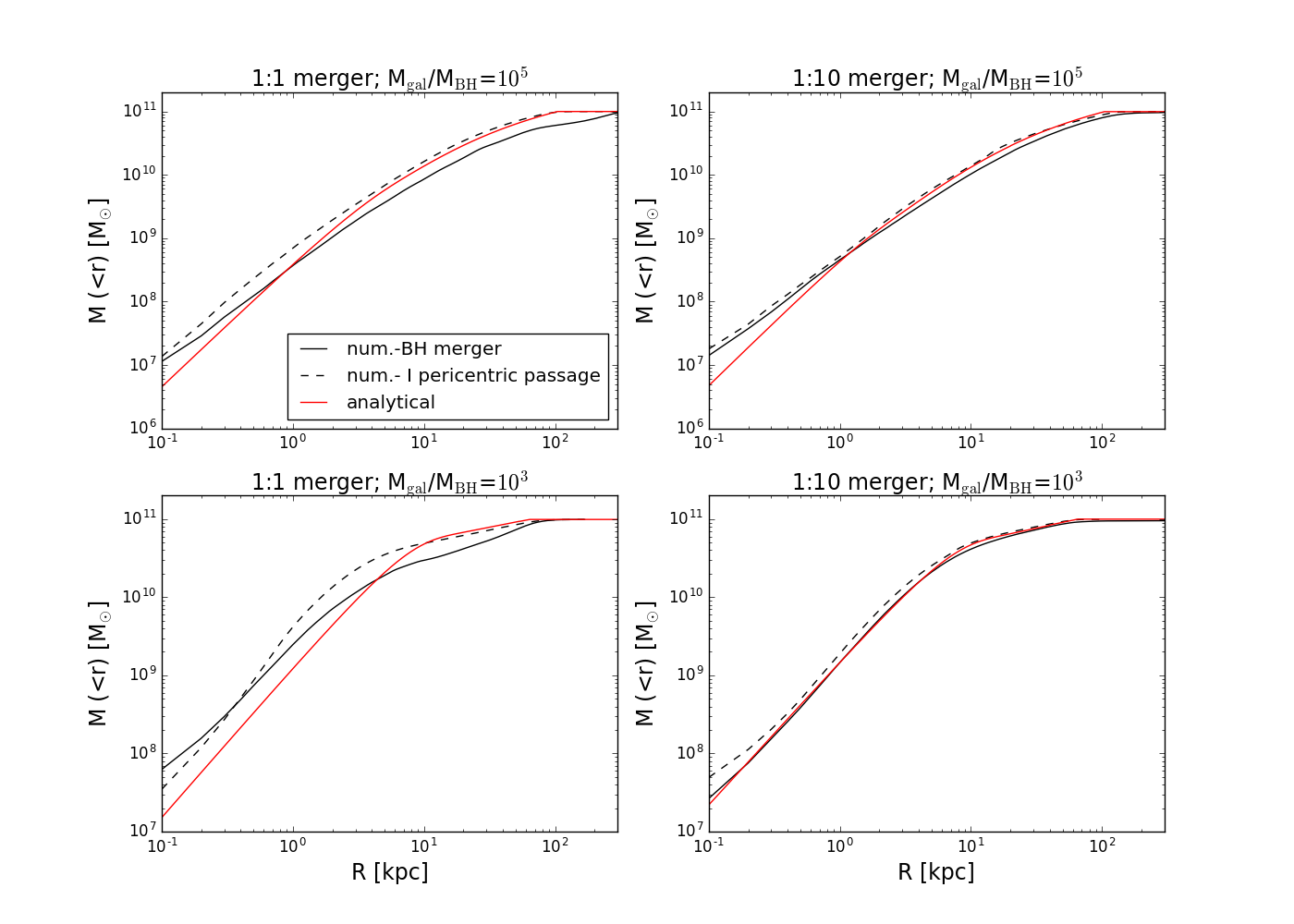}}\\	
	\caption{Total mass profile of numerical and analytical models of merger remnant galaxies. 
      Red lines correspond
      to the analytical models and black lines to numerical models at the first pericentric passage (dashed lines)
      and SMBH merger (solid lines). Panels show major and minor merger remnants of
      galaxies with different central SMBH masses.}
                \label{profil_mase}
	    \end{minipage}

\end{figure*}

If different analytical models are compared with each other, Fig.~\ref{vesc}
shows that galaxies with over-massive SMBHs (${M_{\mathrm{BH}}}=M_{\mathrm{gal}}/10^3$) have
greater escape velocities than galaxies with $M_{\rm{BH}}-M_{\rm{halo}}$ SMBHs  (${M_{\mathrm{BH}}}=M_{\mathrm{gal}}/10^5$).
This is the consequence of two effects:
1) galaxy models with over-massive SMBHs also have massive, extended bulges and thus 
greater density inside central $\sim20$ kpc (mass profiles will be shown on Fig.~\ref{profil_mase}), and 2)  
more massive SMBHs experience greater
drag force due to the dynamical friction (equation~(\ref{fdf})). 
For a given SMBH mass, major and minor merger remnants 
have approximately equal escape velocities. 
Even though major mergers will form an elliptical galaxy and a minor merger product is a disc galaxy,
changes in the total mass profiles between analytical models are negligible.

 \begin{figure*}
 \begin{minipage}{150mm}

	\centering
	\subfloat[Total mass profile of $10^{12}\Msun$ galaxies]{\includegraphics[width=1\columnwidth]{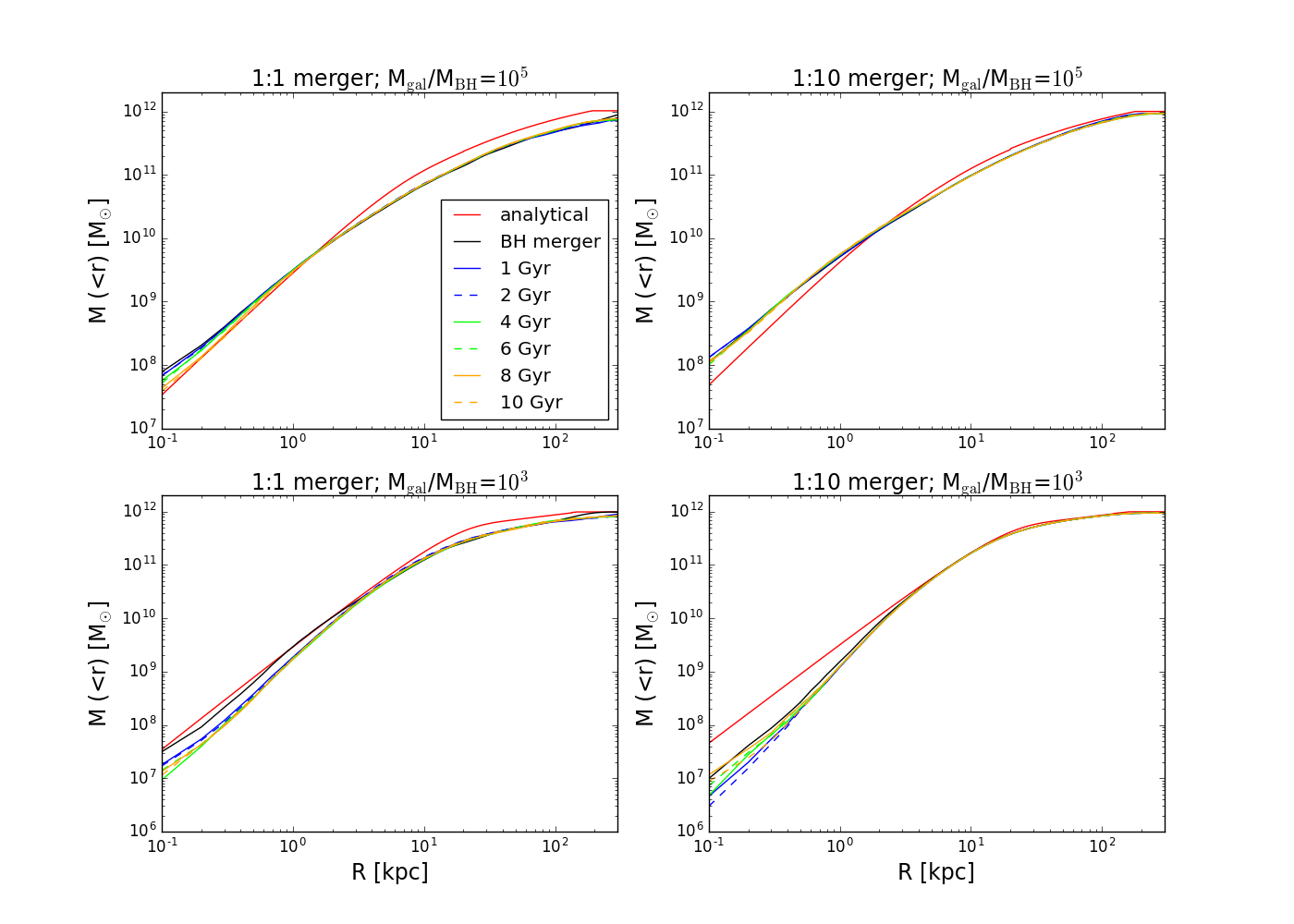}}\\
	\subfloat[Total mass profile of $10^{11}\Msun$ galaxies]{\includegraphics[width=1\columnwidth]{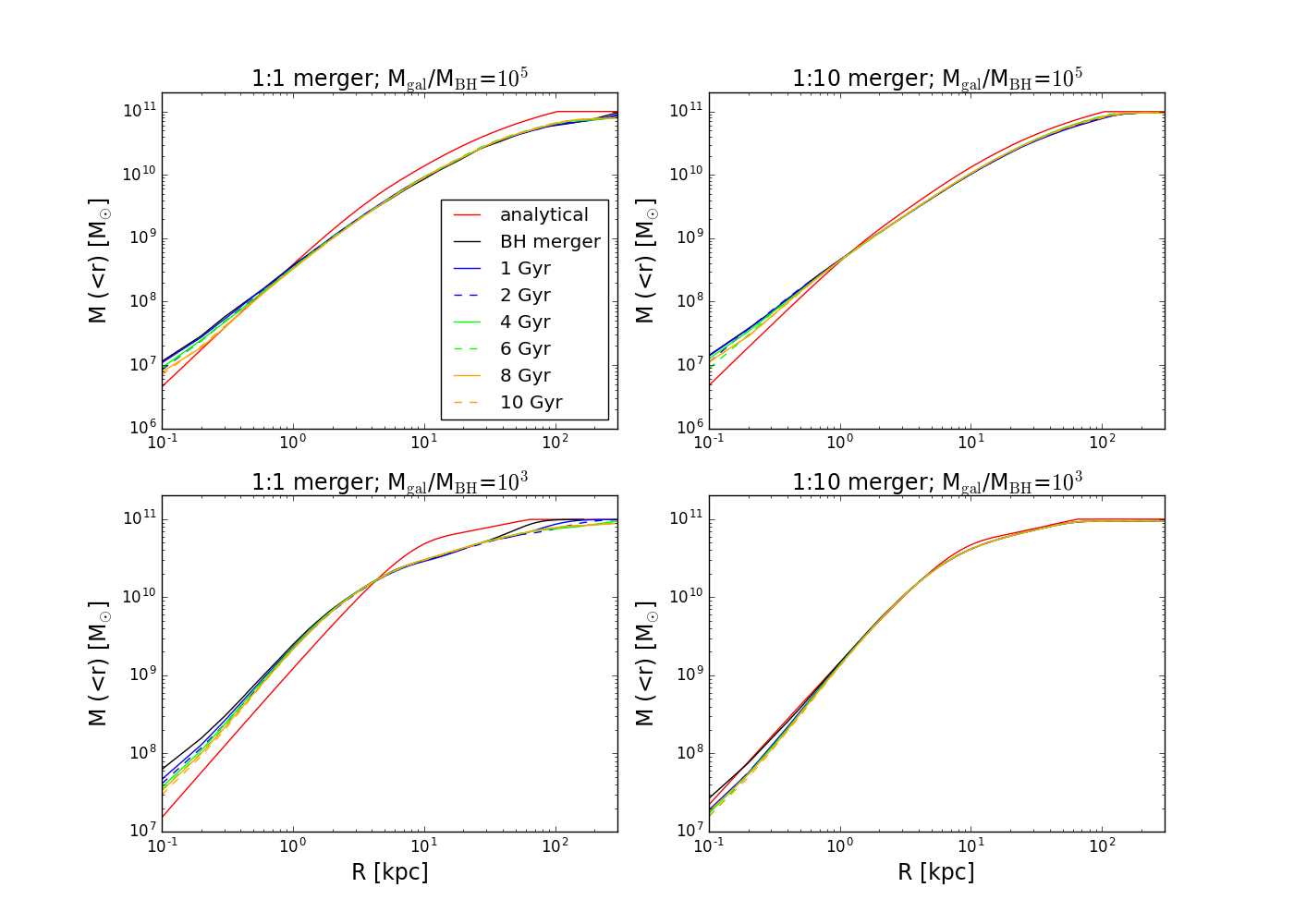}}\\	
	\caption{Evolution of total mass profiles of numerical merger remnant galaxies. Black lines represent
	mass profile at the time of SMBH merger, while blue, green and orange lines show evolution of the total mass
	profiles during the simulation. Red lines represent total mass profile in analytical models.}
                \label{profil_mase_evolution}
	    \end{minipage}

\end{figure*}

%%%%%%%%%%%%%%%%%%%%%%%%%%%%%%%%%%%%%%%%%%%%%%%% numericki sa numerickim
\subsection{Comparison between numerical models}
\label{num_vs_num} 

In the numerical models, mergers lead to redistribution of mass within the merger remnant galaxy.
During the interaction between two galaxies a part of gravitational potential energy is converted 
into the kinetic energy of individual particles and back. This energy conversion is
described by the time dependent virial theorem:

\begin{equation}
\frac{1}{2}\frac{\mathrm{d}^2 I}{{\mathrm{d}}t^2}=2T+V,
\end{equation}
where $T$ and $V$ are kinetic and potential energy of the galaxy, respectively, and
$I$ is the moment of inertia tensor.
Galaxy is virialized, i.e. at equilibrium for $\ddot{I}=0$ and $T=-E$, $V=2E$, where $E$
is the total energy of the galaxy. During galaxy merger, the total energy does not change,
however kinetic and potential energy 
will oscillate around the equilibrium values. 
In such systems energy of individual particles is not conserved. In the process called
violent relaxation \citep{Lynden-Bell}
energy of some particles will decrease 
and those particles will sink to the remnants centre, increasing its central density.
On the other hand, energy of other particles will increase leading to the extended 
mass distribution. Weakly bound particles can escape the host galaxy potential, which results in
mass lost during mergers.
This process is especially efficient during major mergers.
Importance of violent relaxation for major mergers is investigated by \cite{Hilz}.
The authors also showed that in the case of minor mergers violent relaxation influences 
only particles from secondary galaxy and has no significant effect on 
the distribution of the particles in the primary galaxy.

Since SMBH ejection in numerical models occurs in merger remnants 
whose mass and potential are redistributed as a consequence of
the recent galaxy interactions, 
the recoiling SMBH trajectory in numerical models is expected to differ from a
trajectory of a SMBH moving through post-merger galaxy with stable, analytical potential.

When different numerical models are compared mutually, Fig.~\ref{vesc} shows that 
major merger remnants have lower escape velocities than galaxies formed in minor mergers.
Major mergers lead to more significant mass redistribution, which is shown in Fig.~\ref{profil_mase}.
Under the influence of violent relaxation process during major mergers weakly bound particles are
transfered beyond virial radius of the merger remnant. This results in extended mass distribution
with lower mass profile.
Escape velocities in weakened potential of major merger
remnant are $\lesssim25$ per cent  lower compared to minor merger remnants.
Over-massive SMBHs in numerical models also need
larger kick velocities in order to leave their host centre, due to
the presence of massive bulge, and greater influence of dynamical friction force.
Changes in the galaxy mass profile induced by mergers will be discussed in Fig.~\ref{profil_mase}.

%%%%%%%%%%%%%%%%%%%%%%% numericki sa analitickim
\subsection{Comparison between analytical and numerical models}
\label{an_vs_num} 

Finally, if analytical models are compared with numerical models, 
Fig.~\ref{vesc} shows that numerical galaxies have up to 25 per cent lower
escape velocities. 
The greatest difference between analytical and numerical models 
is noticed when escape velocities from major merger remnants with a central
SMBH on $M_{\rm{BH}}-M_{\rm{halo}}$ relation
are compared (open triangles on Fig.~\ref{vesc}).
Those differences are caused by
violent relaxation process during major mergers in numerical models, which 
leads to the decrease of mass in the mass profiles of the numerical remnant galaxies.

Escape velocities from numerical 1:10 merger remnants are
at most 8  per cent lower compared to the analytical models.
Secondary galaxy will experience significant
mass redistribution during minor merger. However, the satellite galaxy has only lesser
contribution to the total potential of the merger remnant.
Changes in the total mass profile caused by violent relaxation process during minor mergers 
are negligible compared to the changes during major mergers.
This leads to the similar escape velocities in numerical and analytical minor merger models.

\subsection{Comparing mass profiles}
\label{mass_profiles} 

In order to further explain differences between escape velocities from different galaxy models, 
we calculate total mass profiles of merger remnants.
Fig. \ref{profil_mase} shows the cumulative mass enclosed within a given radius for different galaxy models.
Major merger remnants are represented in the left panels and minor merger remnants in the right panels.
Galaxies with $M_{\rm{BH}}-M_{\rm{halo}}$ SMBH are shown
in the upper panels and galaxies with over-massive SMBH in the lower panels.
Red solid lines correspond to analytical merger remnant models and black lines to numerical models.
Total mass profile of numerical models is calculated from the snapshot
taken at the time of the first 
pericentric passage (dashed lines) and at the time of SMBH merger (solid lines).

As discussed in Section \ref{num_vs_num} violent relaxation 
during major mergers causes bound particles to become more bound and sink 
to the remnants centre, while weakly bound particles become unbound and escape host galaxy \citep{Lynden-Bell}.
Fig. \ref{profil_mase} shows that at the time of the first pericentric passage, 
total mass of merging galaxies makes a close fit to the analytical mass profile
at radii greater than several kpc (black dashed lines). 
However, at the time of SMBH merger/ejection (black solid lines), 
potential of the major merger remnant has already evolved, 
and when compared to analytical, its mass profile is lower at large radii and higher at small radii. 
A fraction of particles is pushed beyond virial radius which flattens the
mass profile, while some particles sink to the bottom of the 
hosts potential well, increasing central density. 
However, this central mass boost is not sufficient to suppress the SMBH recoil, leading to
the lower escape velocities in numerical major merger remnants.
Thus, the evolving numerical model is a more realistic description of dynamical 
processes in galaxies with merging SMBHs. Static analytical models used in literature, 
overestimate the SMBHs escape velocities, hence, underestimate the number of offset AGNs.

Minor mergers do not induce significant changes in mass profile of the remnant galaxy.
Elliptical galaxies with over-massive SMBH show mass loss in central kpc. This mass loss
is due to interactions of massive central SMBHs with particles in central region.
At large radii deviations from analytical models are negligible.

Fig. \ref{profil_mase_evolution} shows evolution of total mass profiles of numerical merger remnants over 10 Gyr.
Similarly as in Fig. \ref{profil_mase}, different panels correspond to major and minor merger remnants 
of galaxies with different central SMBH masses. 
Black lines show total mass profiles of numerical galaxies at the time of SMBH merger, while blue, 
green and orange lines represent evolution of the post-merger galaxy during the simulation of SMBH ejection.
For a comparison, red lines represent the total mass profile in analytical models.
After the SMBH merger and the ejection of the newly formed SMBH, post-merger galaxies do not show
significant evolution of mass profiles.
Changes in mass profiles are constrained almost entirely to the central kpc of the merger remnant galaxy.
Initial central mass boost induced by major mergers is followed by  
gradual decreasement of central mass profile during the simulation. At large radii 
mass profiles are not affected by this evolution.

\subsection{Recoiling SMBH trajectories}
\label{trajectories} 

\begin{figure}
	\includegraphics[width=\columnwidth]{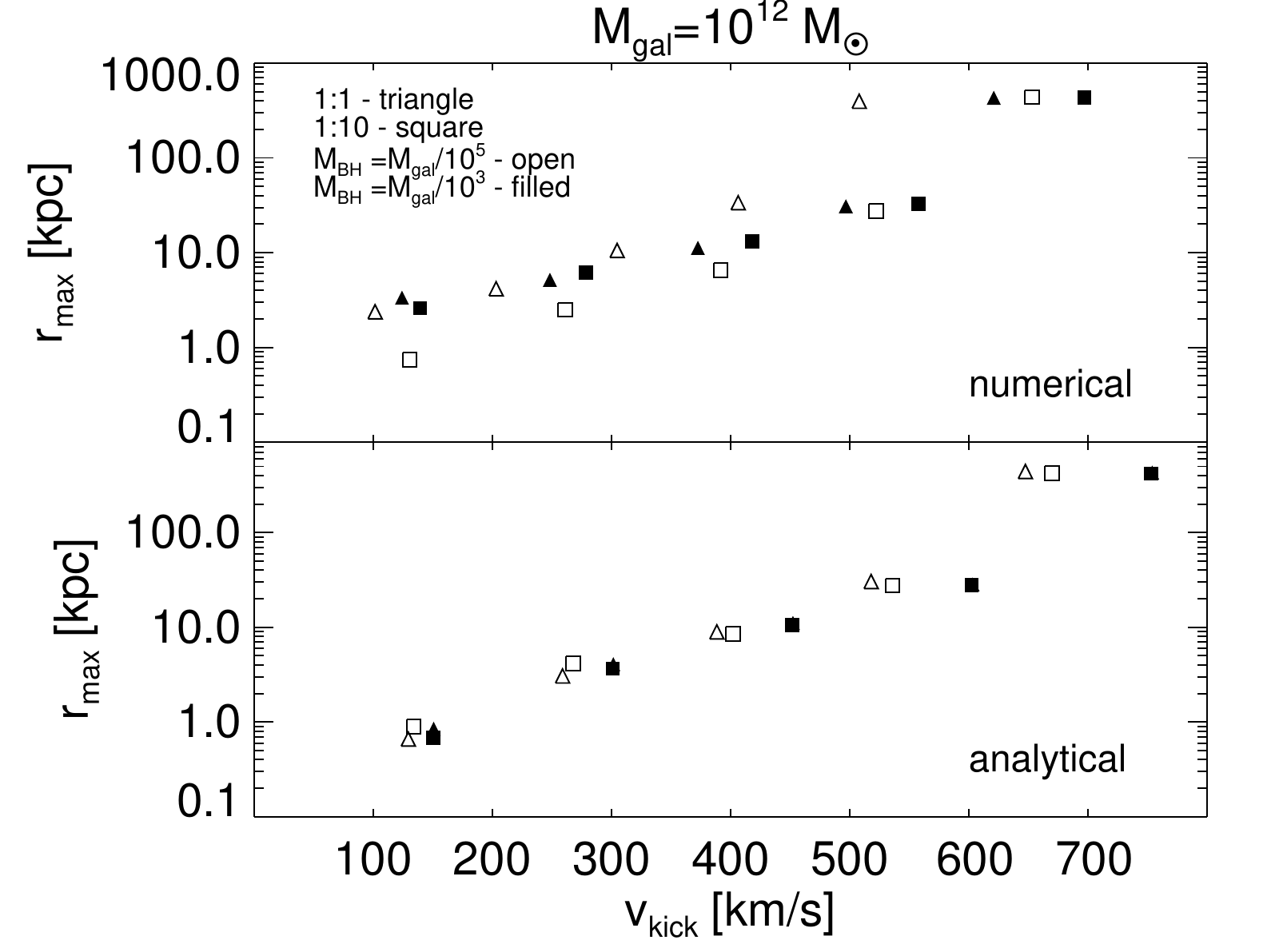}
	\par
	\vspace{0.6cm}
		\includegraphics[width=\columnwidth]{fig7_a.pdf}
 \caption{Maximal separation of a recoiling SMBH from the galaxy centre over a Hubble time, as
 a function of kick velocity. Kick velocities are chosen to represent 
  $v_{\mathrm{kick}}=0.2,0.4,0.6,0.8,1.0\times v_{\mathrm{esc}}$ for each galaxy model.}
    \label{rastojanje}
\end{figure}

Fig. \ref{rastojanje} shows the maximal separation from a galaxy centre reached by recoiled SMBH on a bound orbit
over a Hubble time, for numerical (upper panel) and analytical (lower panel)
models. 
In both analytical and numerical models SMBHs with kick amplitudes 
$v_{\mathrm{kick}}\lesssim300\kms$ for $10^{12}\Msun$ remnants and 
$v_{\mathrm{kick}}\lesssim100\kms$ for $10^{11}\Msun$ remnants, 
stay within central $\sim10$ kpc.
Only SMBHs that receive kick velocities close to the hosts escape velocity 
can reach separation of $r_{\mathrm{max}}>100$ kpc. For these largest separations, 
Fig. \ref{rastojanje} shows that
SMBHs in numerical models reach them with kicks substantially smaller
than in analytical models. For example, SMBH
in numerical major merger remnant of $10^{12}\Msun$ galaxy whose kick 
amplitude is $v_{\mathrm{kick}}\sim500\kms$
will reach approximately the same separation as SMBH with kick amplitude 
$v_{\mathrm{kick}}\sim650\kms$ in analytical model (open triangles on Fig. \ref{rastojanje}).

\begin{figure}
	\includegraphics[width=\columnwidth]{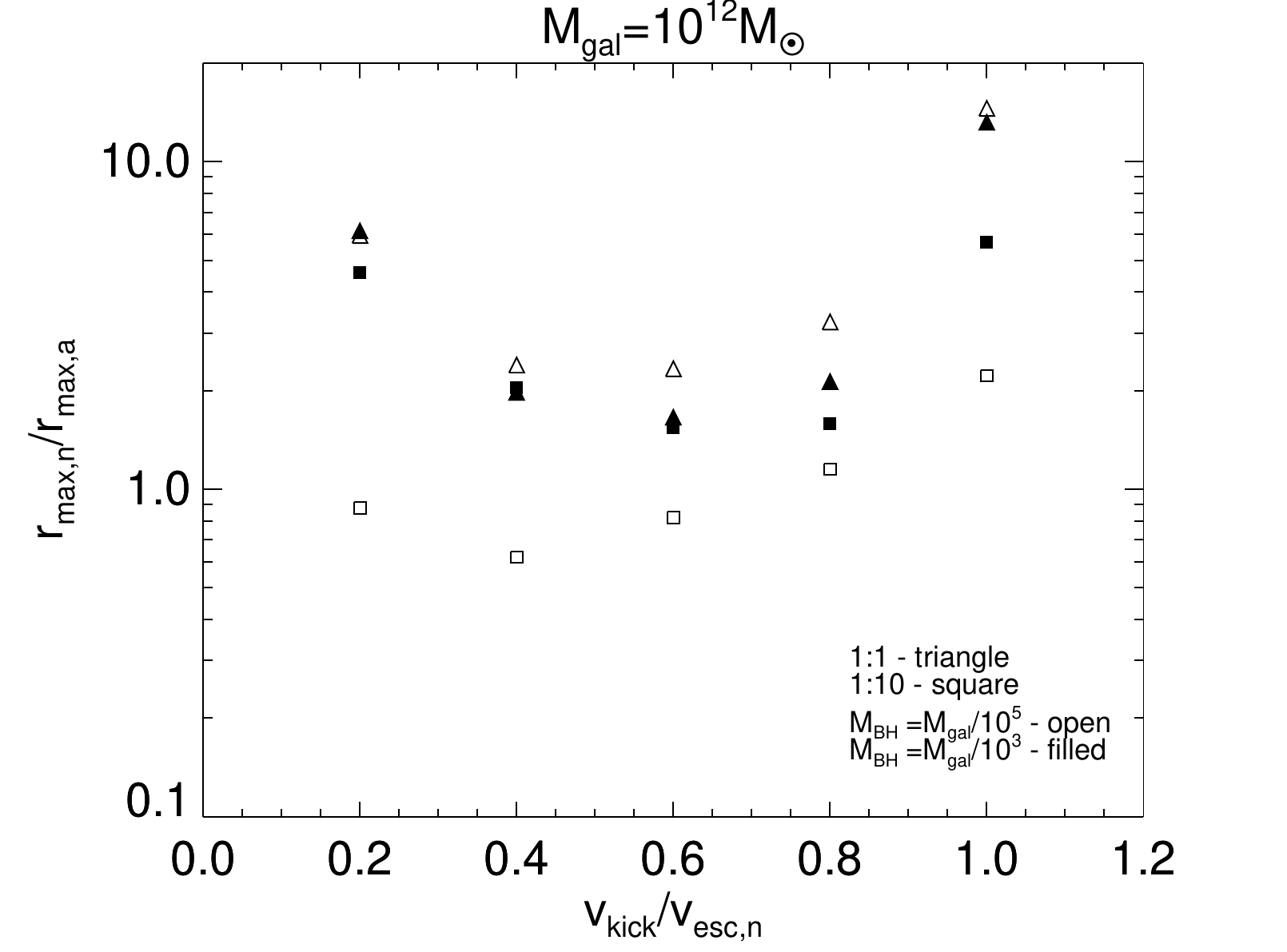}
		\par
	\vspace{0.6cm}
		\includegraphics[width=\columnwidth]{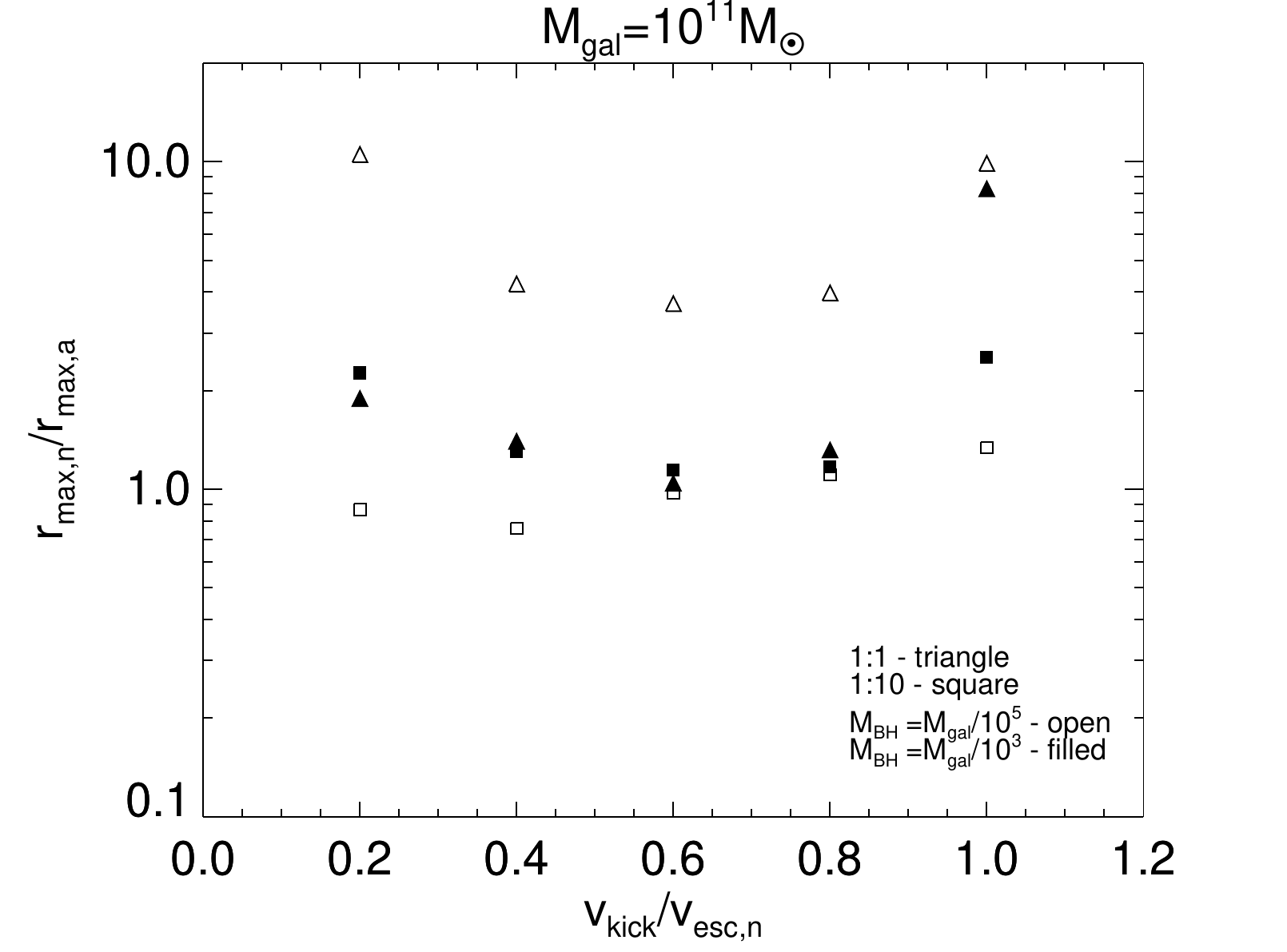}
 \caption{Ratio of the maximal separation of a recoiling SMBH in numerical and analytical models
as a function of $v_{\mathrm{kick}}/v_{\mathrm{esc,n}}$.}
    \label{rmax_razlika}
\end{figure}

In order to further investigate differences between numerical and analytical potentials,
we assign kick velocities to recoiling SMBHs in both analytical and numerical merger 
remnants and compare their distances from galactic centres.
In Fig.~\ref{rmax_razlika} we show ratio of the 
maximal separation from a galaxy centre reached by recoiled SMBH over a
a Hubble time in those two models ($r_{\mathrm{max,n}}/r_{\mathrm{max,a}}$)  
as a function of  $v_{\mathrm{kick}}/v_{\mathrm{esc,n}}$. Here,
 $v_{\mathrm{esc,n}}$ is the escape velocity in numerical post-merger galaxy.
Recoiling SMBHs will reach greater galactocentric distance in numerical models,
which is especially pronounced for major merger remnants.
For low kick velocities, $v_{\mathrm{kick}}/v_{\mathrm{esc,n}}=0.2$,
SMBHs in numerical models will reach several kpc, while in analytical models
SMBHs will stay within central kpc. 
For kick amplitudes equal to the numerical escape velocity,
differences between models become even more significant. 
SMBHs in numerical major merger remnants reach distances up to 10 times larger
than in the analytical models. This result can have large consequences for the predicted distributions of offest AGNs.

\begin{figure}
	\includegraphics[width=\columnwidth]{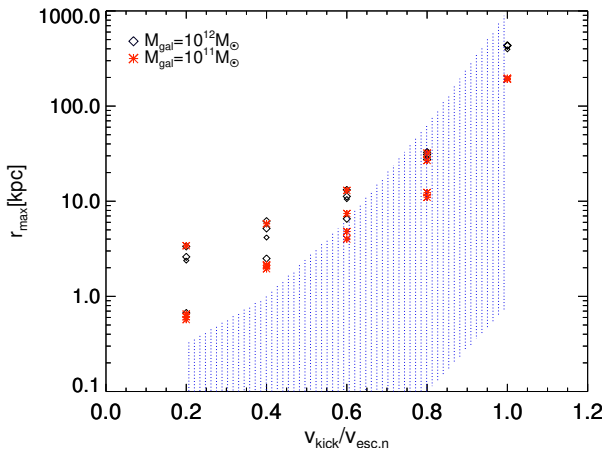}
 \caption{Maximal separation of a recoiling SMBH from the galaxy centre over a Hubble time as
 a function of $v_{\mathrm{kick}}/v_{\mathrm{esc,n}}$, for galaxies with mass 
 $M_{\mathrm{gal}}=10^{12}\Msun$ (diamond symbols) i $M_{\mathrm{gal}}=10^{11}\Msun$
(asterisk symbols). Shaded region represents results obtained by Blecha et al. (2016).}
    \label{rmax_ill}
\end{figure}

Further, we compare our results with the distribution of the ejected SMBHs in
Illustris simulation \citep{blecha16}. 
Fig.~\ref{rmax_ill} shows the maximal separation from a galaxy centre reached by recoiled SMBH
over a Hubble time as
a function of $v_{\mathrm{kick}}/v_{\mathrm{esc,n}}$ for our 
numerical models.
Diamond symbols show galaxies with mass $M_{\mathrm{gal}}=10^{12}\Msun$,
while asterisk symbols represent galaxies with mass $M_{\mathrm{gal}}=10^{11}\Msun$.
Shaded region represents the range of maximal galactocentric separations of recoiled SMBHs
calculated by \cite{blecha16}. The authors assigned SMBH kick velocities calculated from a distribution 
that assumes random pre-merger spin orientation. Spin magnitude 
is chosen from a distribution that peaks at $a\sim0.7$,
where $a$ is dimensionless spin parameters taking values from 
$a=0$ for non-rotating SMBHs to $a\lesssim1$ for SMBHs with maximal rotation.
Distribution of SMBH kick amplitudes resulting from such spin magnitude and spin orientation
distributions peaks at $v_{\mathrm{kick}}\sim300\kms$ and has long tail toward low kick velocities,
although high kicks are also possible.
Fig.~\ref{rmax_ill} shows that the maximal distances from the host centre that SMBHs could reach 
in our numerical models are greater compared to predictions made by \cite{blecha16} for
kick velocities $v_{\mathrm{kick}}/v_{\mathrm{esc,n}}<0.8$.

\cite{blecha16} have shown that, for a random
pre-merger SMBH spin orientation, kick
velocity distribution peaks at $v_{\mathrm{kick}}/v_{\mathrm{esc}}\sim0.3$.
For such kick amplitudes \cite{blecha16} model predicts that recoiling SMBHs will
stay within central kpc, while in numerical models SMBHs can reach $\sim10$ kpc.
SMBHs that reach greater galactocentric distance also spend more time 
on bound orbits outside of the the galactic nucleus, thus for the given $v_{\mathrm{kick}}/v_{\mathrm{esc}}$
ratio numerical models predict more spatially offset AGNs.

Additional difference between analytical and numerical models can be noticed if we compare
escape velocities from the host galaxies with the expected kick velocity distribution for different SMBH spin models,
calculated by \cite{blecha16} (their fig.~1).
If the pre-merger SMBH spins are always aligned with the orbital angular momentum and with each other,
kick velocities are always lower than escape velocities 
of $M_{\mathrm{gal}}=10^{12}\Msun$ galaxies in analytical models. SMBH
could occasionally leave analytical merger remnant with mass $M_{\mathrm{gal}}=10^{11}\Msun$
and central SMBH on $M_{\rm{BH}}-M_{\rm{halo}}$ relation.
Similarly to our analytical results, \cite{blecha16} have found
that SMBHs do not get kick velocities $v_{\mathrm{kick}}/v_{\mathrm{esc}}>0.8$
if SMBH alignment is efficient.
In numerical models, SMBH kick velocities taken from the same 
distribution could be sufficient to eject SMBHs 
from $M_{\mathrm{gal}}=10^{11}\Msun$ galaxies and from the major merger remnant of 
$M_{\mathrm{gal}}=10^{12}\Msun$ galaxy with $M_{\rm{BH}}-M_{\rm{halo}}$ SMBH.
On the other side, if SMBH spins are close to their maximum value and 
randomly oriented prior to merger, SMBHs in both analytical and
numerical models can get kicks $v_{\mathrm{kick}}/v_{\mathrm{esc}}\sim1$.

Kick amplitude also depends on the merging SMBH mass ratio. Mergers of 
SMBHs with similar masses will possibly result with the highest kick amplitudes.
Assuming that SMBH mass ratio scales to the mass ratio of merging
galaxies, recoiling SMBH in the minor merger remnants of $M_{\mathrm{gal}}=10^{12}\Msun$ galaxies may never 
receive kicks with amplitudes similar to the hosts escape velocity.
\cite{micic11} have shown that the interval of possible kick velocities 
for SMBHs with mass ratio 0.1 is $50\lesssim v_{\rm{kick}}\lesssim500\kms$.
In that case, SMBHs in minor merger remnants of massive galaxies will never
get kicks $v_{\mathrm{kick}}/v_{\mathrm{esc}}>0.8$ in numerical models, or
$v_{\mathrm{kick}}/v_{\mathrm{esc}}>0.7$ in analytical models.

Detections of offset AGNs would be helpful in constraining SMBH spin models, however 
observations of such events are challenging. There are several reasons
for that: along with a large sample required for a detection  and limited sensitivity 
or current instruments, offset AGN can be easily mistaken with other phenomena such as
inspiral phase of SMBH mergers or Type IIn supernova. We refer to \cite{blecha19} for a
recent overview of current searches for offset AGN.

Due to the lack of confirmed observations of recoiling SMBHs and offset AGN, our results 
cannot be directly compared with observations.
Instead, results from large cosmological simulations can be used to 
make more concrete predictions about the number of offset 
AGNs in analytical and numerical models.
This can be done semi-analytically by extracting major merger remnant galaxies
with specific masses from the cosmological 
simulation. Those galaxies can be further used to calculate
probability distributions of the recoiling SMBH positions with the respect to galaxy
centre, for both analytical and numerical models. 
Using Monte Carlo method, different kick velocities resulting from different
SMBH spin distributions, can be assigned to SMBHs from the simulation. The final position of the recoiling SMBH
can be then calculated using predictions from analytical and numerical models presented in this paper.
This would yield statistically relevant differences between ejected SMBHs in numerical and analytical models.
However, such analysis is beyond the scope of this paper and it will be the subject of the following research.

\section{Discussion and Conclusions}
\label{discussion}

We have investigated trajectories of recoiling SMBHs in various analytical and numerical models 
of merger remnant galaxies whose components are DMH, disc and bulge.
Considered galaxies have masses of  $M_{\mathrm{gal}}=10^{12}\Msun$ and $M_{\mathrm{gal}}=10^{11}\Msun$
and they host a central SMBH with mass 
${M_{\mathrm{BH}}}=M_{\mathrm{gal}}/{10^5}$ ($M_{\rm{BH}}-M_{\rm{halo}}$ SMBH) or 
${M_{\mathrm{BH}}}=M_{\mathrm{gal}}/{10^3}$ (over-massive SMBH).
Mass of each component is calculated using scaling relations for a given 
mass of the central SMBH and the total galaxy mass. 
We separately model major and minor galaxy mergers in order to investigate if SMBH trajectories
are sensitive to the mass ratio of the progenitor galaxies.

Previous studies of recoiling SMBHs predominantly used static analytical models of merger remnant galaxies in 
order to estimate SMBH escape velocities and the number of offset AGNs.
Our goal was to extend those analytical models of merger remnants and construct 
more realistic numerical
counterparts in order to compare possible differences in recoiling SMBH trajectories. 
First we have shown that analytical and numerical models of isolated progenitor 
galaxies produce the same results for the escape velocity and the 
maximal separation the SMBH could reach over a Hubble time.
Next, we simulate merger of progenitor galaxies and show
that SMBH trajectories in numerical remnants differ from those in 
analytical models.
Initial conditions for numerical galaxy models are generated using 
{\fontfamily{qcr}\selectfont GalactICs} code (\citealt{Kuijken}; \citealt{widrow05}; \citealt{widrow08}),
while galaxy merger simulations are carried out using {\fontfamily{qcr}\selectfont
GADGET-2} code \citep{springel2}.

In our model galaxy escape velocity is defined 
as a SMBH kick velocity necessary for a SMBH to return to its host halo after $\gtrsim10$ Gyr.
We have shown that galaxies with equal total masses have different 
escape velocities due to different mass distributions resulting from the galaxy merger.
Our main results are following:
\begin{enumerate}
\item  Static analytical models overestimate SMBH escape velocities compared to the evolving
numerical models which represent a more realistic description of post-merger galaxies.
Escape velocities in numerical models are up to $\sim25$ per cent lower compared
to analytical models. 
This difference arises from redistribution of mass within post-merger galaxy.
During major mergers violent relaxation process causes bound particles to become more
bound and sink to the remnant centre, and weakly bound particles to become unbound
and escape the host galaxy potential. Escaping particles will make shallower mass profile 
at large radii and reduce escape velocity in major merger remnants.
During minor mergers mass redistribution is not efficient enough
to change the primary galaxy potential and only lesser changes 
in remnants escape velocities are noticed.

\item  In agreement with the previous point, mass ratio of
the progenitor galaxies also influences 
escape velocities of SMBHs in post merger hosts. In numerical models, 
SMBH escape velocities from major merger remnants are up to $\sim25$
per cent lower than escape velocities from numerical minor merger remnants,
for the given SMBH mass.
In analytical models escape velocities from major and minor merger remnants are approximately
equal since analytical models do not account for mass redistribution
induced by violent relaxation process during galaxy mergers.

\item Decreasing modeled SMBH mass from ${M_{\mathrm{BH}}}=M_{\mathrm{gal}}/10^3$ to
${M_{\mathrm{BH}}}=M_{\mathrm{gal}}/10^5$ results in $\lesssim45$ per cent lower
escape velocities in both analytical 
and numerical models. This occurs because galaxies that harbor massive central SMBHs also
have massive bulges ($M_{\mathrm{*}}-M_{\mathrm{BH}}$ relation) which suppress
SMBHs recoils. 
Additional increasement in escape velocities is due to greater gravitational
drag force experienced by over-massive SMBHs (equation (\ref{fdf})).

\item Our model suggests that offset AGNs are more common in numerical models of major merger remnants.
Kick velocities needed to remove a SMBH from the galaxy centre are lower,
which would result in the greater possibility of offset AGNs detection, regardless of
the SMBH spin distribution models.
Differences between numerical and analytical models are significant for both low and
hight kick amplitudes. For
$v_{\mathrm{kick}}/v_{\mathrm{esc,n}}=0.2$, SMBHs in analytical models will stay 
inside central kpc, while in numerical models SMBHs can reach several kpc.
Greater separations are both easier to detect and the return time of a recoiling SMBH is
longer.
For $v_{\mathrm{kick}}/v_{\mathrm{esc,n}}\sim1$,
SMBHs in numerical major merger remnants can reach 
several hundreds of kpc, while in analytical models maximal separation from the host centre in several tens 
of kpc. 
Additionally, the ejected SMBH also carries an accretion disc whose mass
decreases as $v_{\mathrm{kick}}^{2}$ \citep{blecha2008}, so the accretion timescale is longer for SMBHs
with lower kick velocities, which in turn increases possibility of
detection.

\end{enumerate}

Escape velocities from galaxies with mass of $M_{\mathrm{gal}}=10^{12}\Msun$ in numerical models are 
in the range $v_{\mathrm{esc}}\sim500-700\kms$, depending on the SMBH mass and the formation channel 
(minor or major mergers).
If pre-merger SMBH spins are aligned with the orbital angular momentum and with each other 
only rare SMBHs can have kick amplitudes large enough to permanently leave 
major merger remnant with a central SMBH on $M_{\rm{BH}}-M_{\rm{halo}}$ relation.
SMBHs in other models are not expected to get kick velocities $v_{\mathrm{kick}}/v_{\mathrm{esc}}>0.8$.
On the other side, if SMBH spins are randomly oriented even SMBHs with mass $10^{9}\Msun$
could be occasionally ejected from their host centre.
Escape velocities from galaxies with mass of $M_{\mathrm{gal}}=10^{11}\Msun$
are lower, $v_{\mathrm{esc}}\sim170-350\kms$ and greater number of SMBHs are expected
to be found outside their hosts, regardless of SMBH spin parameter and its orientation.
Our numerical models also show that galaxies formed by major mergers have relatively low escape
velocities which could negatively influence merger driven SMBH growth,
since major mergers can trigger episodes of gas accretion and exponential SMBH growth.

As discussed above, more concrete and statistically relevant comparison between numerical and analytical models
would be to calculate probability distributions for SMBH positions in numerical and analytical models, 
using major merger remnants extracted from a cosmological simulations.
This calculation will be the subject of the following research.

\section*{Acknowledgements}

This work was supported by the Ministry of Education, Science and Technological Development of the
Republic of Serbia through project no. 176021, ``Visible and Invisible Matter in Nearby Galaxies:
Theory and Observations.
We thank the anonymous referee for thoughtful
comments that improved our paper.
%%%%%%%%%%%%%%%%%%%%%%%%%%%%%%%%%%%%%%%%%%%%%%%%%%



%\bibliographystyle{mnras}


%%%%%%%%%%%%%%%%%%%%%%%%%%%%%%%%%%%%%%%%%%%%%%%%%%

%%%%%%%%%%%%%%%%% APPENDICES %%%%%%%%%%%%%%%%%%%%%

\appendix

\section{Mass profiles of progenitor galaxies}
\label{A}

 \begin{figure*}
 \begin{minipage}{150mm}

	\centering
	\subfloat[Progenitors of $10^{12}\Msun$ galaxies]{\includegraphics[width=1\columnwidth]{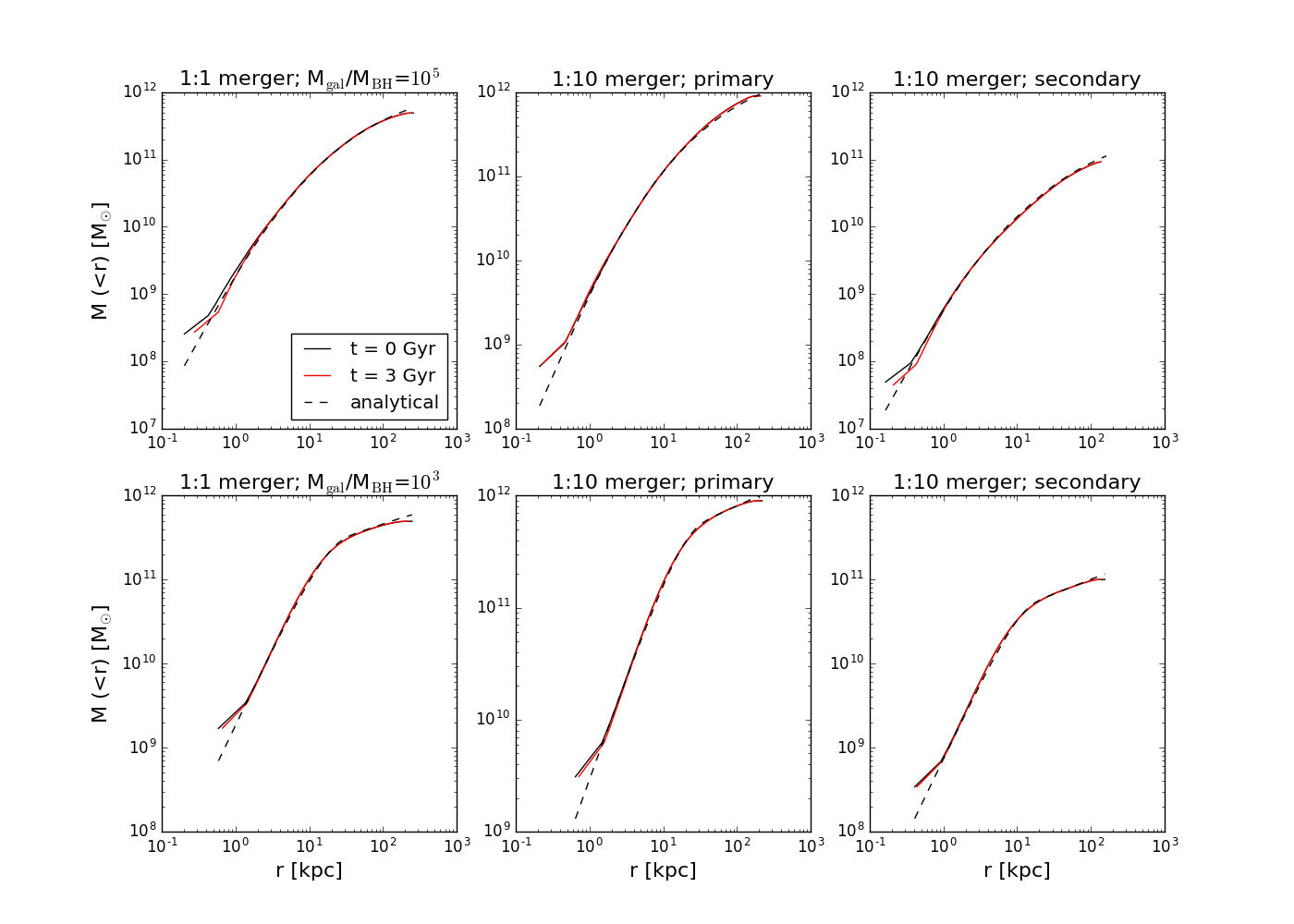}}\\
	\subfloat[Progenitors of $10^{11}\Msun$ galaxies]{\includegraphics[width=1\columnwidth]{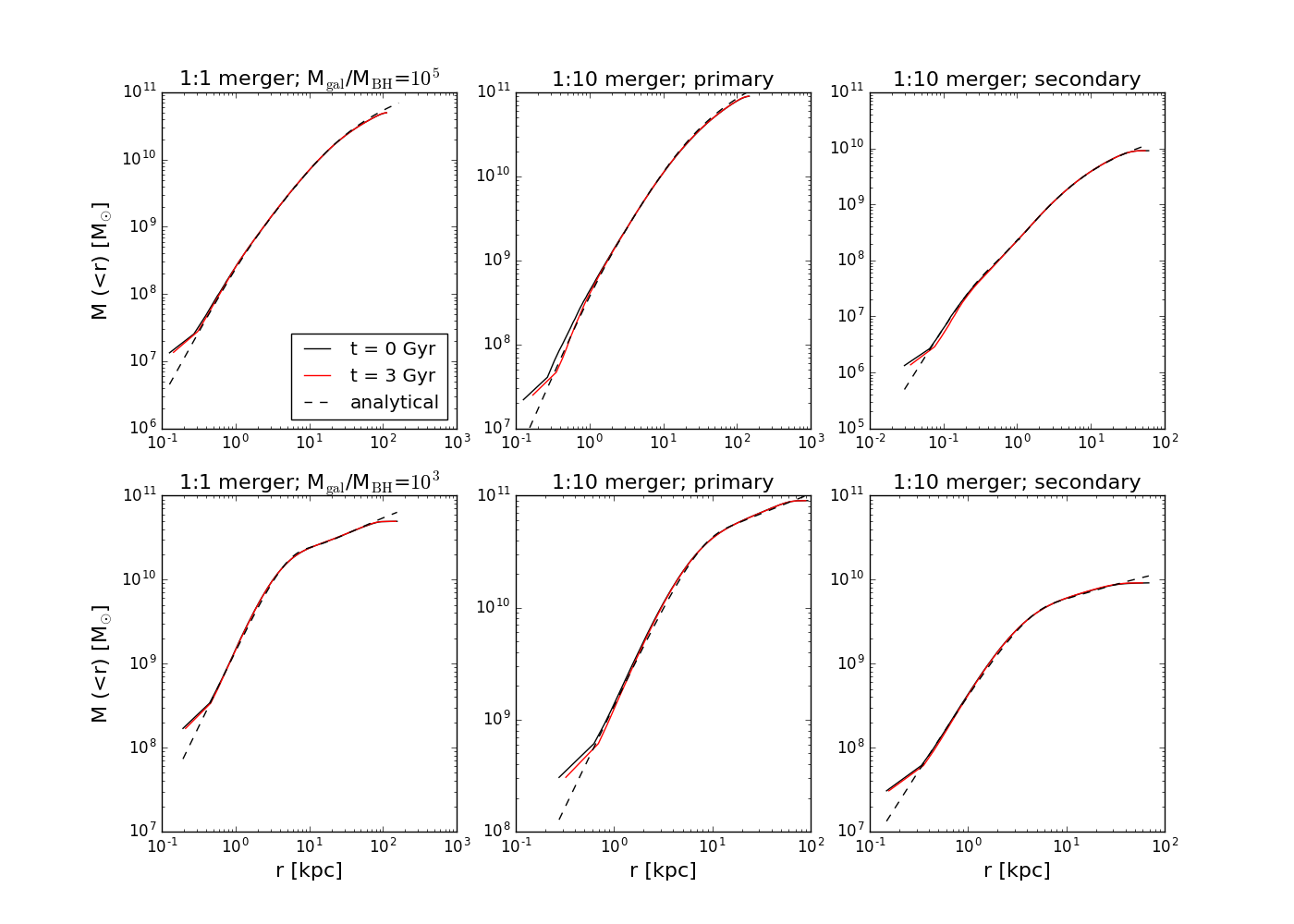}}\\	
	\caption{Total mass profiles in analytical (dashed lines) and numerical (solid lines) galaxy models.
	For numerical models mass profiles of individual galaxy components are shown at the 
	beginning of the simulation and 3 Gyr later. Upper panels represent galaxies with SMBHs on  $M_{\rm{BH}}-M_{\rm{halo}}$
	relation and lower panels galaxies with over-massive BHs.}
	    \end{minipage}

\end{figure*}

%%%%%%%%%%%%%%%%%%%%%%%%%%%%%%%%%%%%%%%%%%%%%%%%%%

% Don't change these lines
\bsp	% typesetting comment
\label{lastpage}
\end{document}